\documentclass{ws-jhde}

\usepackage{epstopdf}

\begin{document}

Preprint of an article published in Journal of Hyperbolic Differential Equations Vol. 16, No. 4 (2019) 1Ц25 \copyright World Scientific Publishing Company

DOI: 10.1142/S0219891619500243

\markboth{A.M. Blokhin, D.L. Tkachev}
{Stability of Poiseuille-type flows for an MHD model of an incompressible polymeric fluid}

%
%

\title{Stability of Poiseuille-type flows for an MHD model of an incompressible polymeric fluid}

\author{A.M. Blokhin}

\address{Sobolev Institute of Mathematics, Koptyug av., 4\\
Novosibirsk, 630090, Russia\\
Novosibirsk State University, Pirogova st., 1\\
Novosibirsk, 630090, Russia}
\email{blokhin@math.nsc.ru}

\author{D.L. Tkachev}

\address{Sobolev Institute of Mathematics, Koptyug av., 4\\
Novosibirsk, 630090, Russia\\
Novosibirsk State University, Pirogova st., 1\\
Novosibirsk, 630090, Russia}
\email{tkachev@math.nsc.ru}

\maketitle

\begin{history}
\received{(Day Mth. Year)}
\revised{(Day Mth. Year)}
\comby{[editor]}
\end{history}

\begin{abstract}
{\bfseries Abstract.}\quad We study a new rheological model describing flows of melts and solutions of incompressible viscoelastic polymeric media in an external uniform magnetic field in the presence of a temperature drop and a conduction current. We find an asymptotic representation of the spectrum of the linear problem resulting from the linearization of the initial boundary value problem in an infinite plane channel about a Poiseuille-type flow. For this Poiseuille-type flow we also find the parameter domain of linear Lyapunov's stability.
\end{abstract}


\section{Introduction}

In this work we study a generalization of the structurally phenomenological Pokrovski--Vinogradov model describing flows of melts and solutions of incompressible viscoelastic polymeric media to the non-isothermal case under the influence of magnetic field. In the Pokrovski--Vinogradov model, the polymeric medium is considered as a suspension of polymer macromolecules that move in an anisotropic fluid produced, for example, by a solvent and other macromolecules. The influence of environment on a real macromolecule is modeled by the action on a linear chain of brownian particles, each of which represents a large enough part of the macromolecule. Brownian particles often called ``beads'' are connected to each other by elastic forces called ``springs''. In the case of slow motions the macromolecule is modeled as a chain of two particles called ``dumbbell''.

The physical representation of linear polymer flows described above results in the formulation of the Pokrovski--Vinogradov rheological model \cite{20}--\cite{1}:
\begin{equation}
\rho(\frac{\partial}{\partial t}v_{i} + v_{k}\frac{\partial}{\partial x_{k}}v_{i}) = \frac{\partial}{\partial x_{k}}\sigma_{ik}, \quad \frac{\partial v_{i}}{\partial x_{i}} = 0, \label{0.1}
\end{equation}
\begin{equation}
\sigma_{ik} = -p\delta_{ik} + 3 \frac{\eta_{0}}{\tau_{0}}a_{ik}, \label{0.2}
\end{equation}
\begin{equation}
\frac{d}{dt}a_{ik} - v_{ij}a_{jk} - v_{kj}a_{ji} + \frac{1 + (k - \beta)I}{\tau_{0}}a_{ik} = \frac{2}{3} \gamma_{ik} - \frac{3\beta}{\tau_{0}}a_{ij}a_{jk}, \label{0.3}
\end{equation}
where $\sigma_{ik}$ is the stress tensor, $p$ is the hydrostatic pressure, $\eta_{0}$, $\tau_{0}$ are the initial values of the shear viscosity and the relaxation time for the viscoelastic component, $v_{ij}$ is the tensor of the velocity gradient, $a_{ik}$ is the symmetrical tensor of additional stresses of second rank, $I = a_{11} + a_{22} + a_{33}$ is the first invariant of the tensor of additional stresses, $\gamma_{ik} = \frac{v_{ik} + v_{ki}}{2}$ is the symmetrized tensor of the velocity gradient, $k$ and $\beta$ are the phenomenological parameters taking into account the shape and the size of the coiled molecule in the dynamics equations of the polymer macromolecule, $\rho$ is the polymer density, $v_{i}$ is the $i$-th velocity component.

Structurally, the model consists of the incompressibility and motion equations \eqref{0.1} as well as the rheological relations \eqref{0.2}, \eqref{0.3} connecting kinematic characteristics of the flow and its inner thermodynamic parameters.

Some generalizations of model \eqref{0.1} -- \eqref{0.3} provide good results in numerical simulations  for viscosymetric flows \cite{22}. For example, such a generalization is a model for which in equation \eqref{0.2} we add a term taking into account the so-called shear viscosity, and the parameter  $\beta$ is additionally dependent on the first invariant of the anisotropy tensor. Therefore,
we may believe  that modifications of the basic Polrovski--Vinogradov model could be useful for modeling the polymer motion in complex deformation conditions, e.g., for stationary and non-stationary flows in circular channels, flows in channels with a fast change of sectional area and flows with a free boundary. An important feature of such flows is their two- and three dimensional character.

In this work, we consider one of such generalizations that takes into account the influence of the heat and the magnetic field on the  polymeric fluid motion (see Sect. 1 for more details). Our main interest is the analogue of the well-known shear flow in an infinite channel that is the Poiseuille flow. It turns out that in our case this flow has a number of features. For example, computations show that for some values of parameters the velocity profile is stretched in the direction opposite to the forces of pressure (see Sect. 2).

Our main results are given in Sect. 4. Firstly, we get an asymptotic representation of the spectrum of the  problem linearized about the the chosen basic solution which is the Poiseuille-type flow. Secondly, as the result we get a condition whose fulfilment guarantees that the basic solution is asymptotically stable by Lyapunov in the chosen class of perturbations periodic with respect the variable changing along the infinite plane channel. The last section is devoted to the proof of the theorems formulated in Sect. 4.

It should be noted that for the case of viscous fluid there is known result of Krylov about the linear Lyapunov's instability  of the Poiseuille flow for large enough Reynolds numbers (see \cite{23}) confirming Heisenberg's hypothesis \cite{24} (a refinement of this result was obtained in \cite{25}).

\section{Nonlinear model of the polymeric fluid flow in a plane channel under the presence of an external magnetic field}

Using the results from \cite{1}, \cite{2}--\cite{5} and \cite{6}, let us formulate a mathematical model describing magnetohydrodynamic flows of an incompressible polymeric fluid for which, as in \cite{7}, in the equation for the inner energy we introduce some dissipative terms. In a dimensionless form this model reads (we keep the notations from \cite{6}):
\begin{equation}
div\vec{\boldsymbol{u}} = u_{x} + v_{y} = 0, \label{1.1}
\end{equation}
\begin{equation}
div\vec{H} = L_{x} + M_{y} = 0, \label{1.2}
\end{equation}
\begin{equation}
\frac{d\vec{\boldsymbol{u}}}{dt} + \nabla P = div(Z\Pi) + \sigma_{m}(\vec{H}, \nabla)\vec{H} + Gr(Z-1)\begin{pmatrix}0\\1\end{pmatrix}, \label{1.3}
\end{equation}
\begin{equation}
\frac{da_{11}}{dt} - 2A_{1}u_{x} - 2a_{12}u_{y} + L_{11} = 0, \label{1.4}
\end{equation}
\begin{equation}
\frac{da_{22}}{dt} - 2A_{2}v_{y} - 2a_{12}v_{x} + L_{22} = 0, \label{1.5}
\end{equation}
\begin{equation}
\frac{da_{12}}{dt} - A_{1}v_{x} - A_{2}u_{y} + \frac{\widetilde{K}_{I}a_{12}}{\bar{\tau}_{0}(Z)} = 0, \label{1.6}
\end{equation}
\begin{equation}
\frac{dZ}{dt} = \frac{1}{Pr}\Delta_{x,y}Z + \frac{A_{r}}{Pr}ZD + \frac{A_{m}}{Pr}\sigma_{m}D_{m}, \label{1.7}
\end{equation}
\begin{equation}
\frac{d\vec{H}}{dt} - (\vec{H},\nabla)\vec{\boldsymbol{u}} - b_{m}\Delta_{x,y}\vec{H} = 0. \label{1.8}
\end{equation}
where $t$ is the time, $u$, $v$ and $L$, $1+M$ are the components of the velocity vector $\vec{\boldsymbol{u}}$ and the magnetic field $\vec{H}$ respectively in the Cartesian coordinate system $x,y$; $$P = \rho + \sigma_{m}\frac{L + (1+ M)}{2},$$ $\rho$ is the pressure;
$a_{11}$, $a_{22}$, $a_{12}$ are the components of the symmetrical anisotropy tensor of second rank; $$\Pi = \frac{1}{Re}(a_{ij}),\quad i,j = 1,2;\quad L_{ii} = \frac{K_{I}a_{ii} + \beta(a_{ii}^{2} + a_{12}^{2})}{\bar{\tau}_{0}(Z)},\quad i = 1,2;$$
$$K_{I} = W^{-1} + \frac{\bar{k}}{3}I,\quad \bar{k} = k - \beta ;$$ $I = a_{11} + a_{22}$ is the first invariant of the anisotropy tensor; $k$, $\beta$ ($0 < \beta < 1$) are the phenomenological parameters of the rheological model (see \cite{1});
$A_{i} = W^{-1} + a_{ii}$, $i = 1,2$; $Z = \frac{T}{T_{0}}$; $T$ is the temperature, $T_{0}$ is an average temperature (room temperature; we will further assume that $T_{0} = 300$ K);\\
$\widetilde{K}_{I} = K_{I} + \beta I$; $\bar{\tau}_{0}(Z) = \frac{1}{ZJ(Z)}$, $J(Z) = \exp\{\bar{E}_{A}\frac{Z-1}{Z}\}$,\\
$\bar{E}_{A} = \frac{E_{A}}{T_{0}}$, $E_{A}$ is the activation energy;\\
$Re = \frac{\rho u_{H}l}{\eta_{0}^{*}}$ is the Reynolds number;\\
$W = \frac{\tau_{0}^{*}u_{H}}{l}$ is the Weissenberg number;\\
$Gr = \frac{Ra}{Pr}$ is the Grasshoff number;\\
$Pr = \frac{lu_{H}\rho c_{v}}{\varepsilon} = \frac{c_{v}\eta_{0}^{*}Re}{\varepsilon}$ is the Prandtl number;\\
$Ra = \frac{lbgT_{0}Pr}{u_{H}^{2}}$ is the Rayleigh number;\\
$A_{r} = \frac{\alpha u_{H}^{2}Pr}{ReT_{0}c_{v}} = \frac{\alpha u_{H}^{2}\eta_{0}^{*}}{T_{0}\varepsilon}$, $A_{m} = \frac{\alpha_{m} u_{H}^{2}Pr}{T_{0}c_{v}}$;\\
$D = a_{11}u_{x} + (v_{x} + u_{y})a_{12} + a_{22}v_{y}$;\\
$D_{m} = L^{2}u_{x} + L(1 + M)(v_{x} + u_{y}) + (1 + M)^{2}v_{y}$;\\
$\rho(= const)$ is the medium density;\\
$\varepsilon$ is the coefficient of thermal conductivity  of the polymeric fluid;\\
$b$ is the coefficient thermal expansion of the polymeric fluid;\\
$g$ is the gravity constant;\\
$\eta_{0}^{*}$, $\tau_{0}^{*}$ are the initial values for the shear viscosity and the relaxation time for the room temperature $T_{0}$ (see \cite{1,6});\\
$l$ is the characteristic length, $u_{H}$ is the characteristic velocity;\\
$\sigma_{m} = \frac{\mu\mu_{0}H_{0}^{2}}{\rho u_{H}^{2}}$ is the magnetic pressure coefficient;\\
$b_{m} = \frac{1}{Re_{m}}$, $Re_{m} = \sigma_{m}\mu\mu_{0}u_{H}l$ is the magnetic Reynolds number;\\
$\mu_{0}$ is the magnetic penetration in vacuum;\\
$\mu$ is the magnetic penetration of the polymeric fluid;\\
$\sigma$ is the electrical conductivity of the medium;\\
$\alpha$ is the thermal equivalent of work (see \cite{8});\\
$\alpha_{m}$ is the magnetothermal equivalent of work;\\
$c_{v}$ is the heat capacity;\\
$\frac{d}{dt} = \frac{\partial}{\partial t} + (\vec{\boldsymbol{u}}, \nabla) = \frac{\partial}{\partial t} + u\frac{\partial}{\partial x} + v\frac{\partial}{\partial y}$,\\
$\Delta_{x,y} = \frac{\partial^{2}}{\partial x^{2}} + \frac{\partial^{2}}{\partial y^{2}}$ is the Laplace operator.\\
The variables $t$, $x$, $y$, $u$, $v$, $p$, $a_{11}$, $a_{22}$, $a_{12}$, $L$, $M$ in system \eqref{1.1}--\eqref{1.8} correspond to the following values: $\frac{l}{u_{H}}$, $l$, $u_{H}$, $\rho u_{H}^{2}$, $\frac{W}{3}$, $H_{0}$, where $H_{0}$ is the characteristic magnitude of the magnetic field  (see fig. 1).

\begin{figure}[ph]
\centerline{\psfig{file=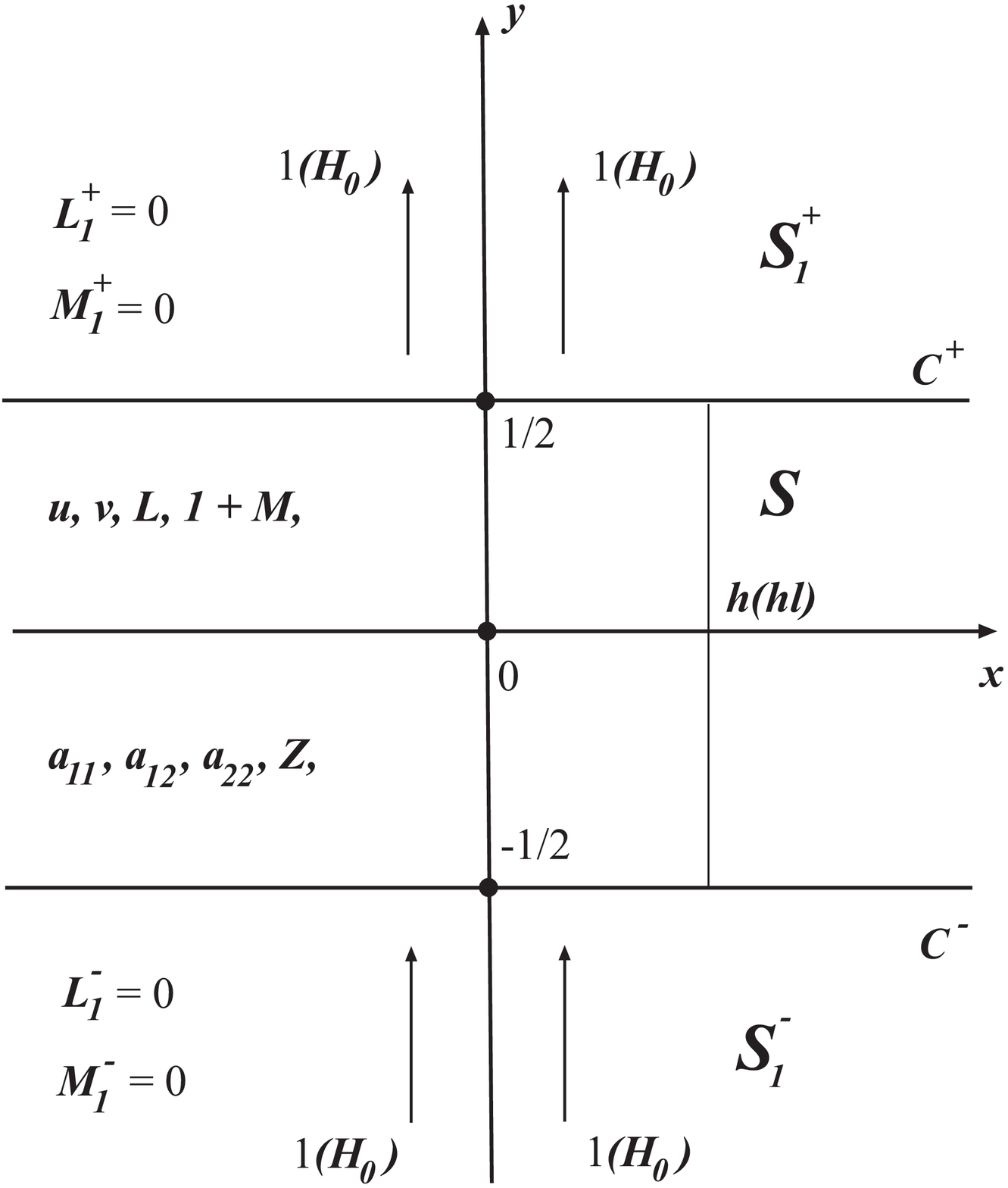,width=1.5in}} 
\vspace*{8pt}
\caption{Plane channel}
\end{figure}

\begin{remark}
The magnetohydrodynamic equations \eqref{1.1} -- \eqref{1.8} are derived with the use of the Maxwell equations (see \cite{2,4}). The magnetic induction vector $\vec{B}$ is represented as
\begin{equation}
\vec{B} = \mu\mu_{0}\vec{H} = (1 + \chi)\mu_{0}\vec{H}, \label{1.9}
\end{equation}
where $\chi$ is the magnetic susceptibility, and (see \cite{9,10}) $\chi = \frac{\chi_{0}}{Z}$, $\chi_{0}$ is the magnetic susceptibility for the room temperature $T_{0}$(= 300 K). We will further assume that for the polymeric fluid $\mu =1$ ($\chi_{0} = 0$).
\end{remark}

Our main problem is the problem of finding solutions to the mathematical model \eqref{1.1} -- \eqref{1.8} describing magnetohydrodynamic flows of an incompressible polymeric fluid in a plane channel with the depth $1(l)$ and bounded by horizontal walls which are the electrodes $C^{+}$ and $C^{-}$ along which we have  electric currents with the current strength of $J^{+}$ of $J^{-}$ respectively (see Fig. 1).

The channel is placed in a uniform external magnetic field with the components $L = 0$ and $M = 0$, i.e., $1 + M = 1 (H_{0})$.

The domains $S_{1}^{+}$ and $S_{1}^{-}$ external for the channel  are magnets with the magnetic susceptibilities $\chi_{1}^{+}$ and $\chi_{1}^{-}$. On the walls of the channel the following boundary conditions hold:
\begin{equation*}
y = \pm\frac{1}{2}: \quad \vec{\boldsymbol{u}} = 0 \quad \mbox{(no-slip condition)},
\end{equation*}
\begin{equation}
y = \frac{1}{2}: \quad Z = 1,  \label{1.10}
\end{equation}
 i.e., the temperature $T = T_{0}$ in the domain $S_{1}^{+}$ and on the electrode $C^{+}$;
\[
y = -\frac{1}{2}: Z = 1 + \bar{\theta}, \bar{\theta} = \frac{\theta}{T_{0}}, \theta = T - T_{0},
\]
i.e., for $\bar{\theta} > 0$ there is  heating from below ($T$ is the temperature in the domain $S_{1}^{-}$ and on the electrode $C^{-}$), and for $\bar{\theta} < 0$ there is heating from above.

Due to \eqref{1.9} and \eqref{1.10},
\begin{equation}
\chi_{1}^{+} = \chi_{0}^{+}, \quad \chi_{1}^{-} = \frac{\chi_{0}^{-}}{1 + \bar{\theta}}. \label{1.11}
\end{equation}

\begin{remark}
We will further assume that either $k = \beta$ or $k = 1.2\beta$. Note that, as follows from \cite{1}, for $k = 1.2\beta$  we get results which have a best correspondence to experiments data.
\end{remark}
\begin{remark}
The boundaries $C^{+}$ and $C^{-}$ are the boundaries between two isotropic magnets. Consequently, on the boundaries $C^{+}$ and $C^{-}$ the following boundary conditions hold (see \cite{9,11}):
\begin{equation}
\begin{array}{l}
y = \frac{1}{2}\,(C^{+}): \quad L = -J^{+}, \quad M = \chi_{0}^{+},\\
y = -\frac{1}{2}\,(C^{-}): \quad L = -J^{-}, \quad M = \frac{\chi_{0}^{-}}{1 + \bar{\theta}}.
\end{array} \label{1.12}
\end{equation}
\end{remark}

In this work we consider a Poiseuile-type flow as the basic flow of an incompressible polymeric fluid which is one of the stationary solutions for the Navier--Stokes equations. If instead of the no-slip condition on the boundary $y = \frac{1}{2}$ $\vec{\boldsymbol{u}} = 0$ we set $u = 1$, $v = 0$, then we get a Couette-type flow (see \cite{3,8}).

\begin{remark}
We again note that, unlike what we do in \cite{6}, \cite{12}--\cite{14}, equation \eqref{1.7} for the inner energy (heat influx) includes dissipative terms characterizing the heat flow arising for nonzero velocity gradients. The existence of such dissipative terms results, in particular, in nonisothermality and $\bar{\theta} = 0$ (that is correct because the thermal influx exists due to the work of the anisotropy tensor components $\hat{a}_{11}$, $\hat{a}_{12}$ and $\hat{a}_{22}$).
\end{remark}

For the model from \cite{6} one has, on the contrary, that  $\tau_{0}(\hat{Z}) = 1$ for $\bar{\theta} = 0$ and $\hat{Z} = 1$, i.e., there appears an isothermic process (here and below all the hat values stay for the basic solution).

\section{Magnetohydrodynamic stationary flows in a plane channel. Poiseuille-type flows}

Let us denote $\vec{U}(t,x,y) = (u,v,a_{11},a_{12},a_{22},Z,L,M)^{T}$. We will seek a partial solution of system \eqref{1.1}--\eqref{1.8} in the special form
\begin{equation}
\left\{
\begin{array}{l}
\vec{U}(t,x,y) = \vec{U}(y),\\
p(t,x,y) = \hat{P}(y) + \hat{P}_{0} - \hat{A}x
\end{array} \right. \label{2.1}
\end{equation}
corresponding to the stationary flow of an incompressible polymeric fluid in an infinite plane channel (see Fig. 1) under the effect of a constant pressure drop along the channel axis $y = 0$. Here $\vec{U}(y) = (\hat{u}(y), 0, \hat{a}_{11}(y), \hat{a}_{12}(y), \hat{a}_{22}(y), \hat{Z}(y), \hat{L}(y), \hat{M}(y))^{T}$, $\hat{P}(y)$ ($\hat{P}(0) = 0$) is some function that requires a further definition, $\hat{p}_{0}$ is the pressure on the channel axis for $y = 0$, $x = 0$, $\hat{A} = \frac{\widehat{\Delta p}}{\rho u_{H}^{2}h}$, $-\frac{\widehat{\Delta p}}{\rho u_{H}^{2}h}$ is the dimensionless pressure drop on the segment $h$, where the dimensional value $\widehat{\Delta p} > 0$ (see Fig. 1).

For finding the functions $\hat{u}(y)$, $\hat{a}_{11}(y)$, $\hat{a}_{12}(y)$, $\hat{a}_{22}(y)$, $\hat{Z}(y)$, $\hat{L}(y)$, $\hat{M}(y)$ and $\hat{P}(y)$ from \eqref{1.1} -- \eqref{1.8}, \eqref{1.10}, \eqref{1.12} we get the following equations:
\begin{equation}
\left(\hat{Z}\hat{a}_{12} + (1 + \hat{\lambda})\sigma_{m}Re\hat{L}\right)' = -\hat{D}, \quad \hat{D} = Re\hat{A}, \label{2.2}
\end{equation}
\begin{equation}
\left(\hat{P} + \sigma_{m}\frac{\hat{L}^{2}}{2} - \frac{\hat{Z}\hat{a}_{22}}{Re}\right)' = Gr(\hat{Z} - 1), \quad \hat{P}(0) = 0, \label{2.3}
\end{equation}
\begin{equation}
u' = \frac{\widetilde{K}_{\hat{I}}J(\hat{Z})\hat{Z}\hat{a}_{12}}{\hat{A}_{2}}, \quad \hat{u}(\frac{1}{2}) = \hat{u}(-\frac{1}{2}) = 0, \quad \hat{I} = \hat{a}_{11} + \hat{a}_{22}, \label{2.4}
\end{equation}
\begin{equation}
K_{\hat{I}}\hat{a}_{22} + \beta(\hat{g} + \hat{a}_{22}^{2}) = 0, \quad \hat{g} = \hat{a}_{12}^{2}, \label{2.5}
\end{equation}
\begin{equation}
K_{\hat{I}}\hat{a}_{11} + \beta(\hat{g} + \hat{a}_{11}^{2}) - 2\hat{g}\frac{K_{\hat{I}}}{\hat{A}_{2}} = 0, \label{2.6}
\end{equation}
\begin{equation}
\hat{Z}'' + (A_{r}\hat{Z}\hat{a}_{12} + A_{m}\sigma_{m}(1 + \hat{\lambda})\hat{L})\hat{u}' = 0, \, \hat{Z}(\frac{1}{2}) = 1,\, \hat{Z}(-\frac{1}{2}) = 1 + \bar{\theta}, \label{2.7}
\end{equation}
\begin{equation}
b_{m}\hat{L}'' + (1 + \hat{\lambda})\hat{u}' = 0, \quad \hat{L}(\pm\frac{1}{2}) = -J^{\pm}, \label{2.8}
\end{equation}
\begin{equation}
\hat{M}'' = 0, \quad \hat{M}' = 0,\, \mbox{i.e. } \hat{M} = \hat{\lambda} = \chi_{0}^{+} = \frac{\chi_{0}^{-}}{1 + \bar{\theta}} = const \, (\mbox{see \eqref{1.12}}). \label{2.9}
\end{equation}
From \eqref{2.2} we get:
\begin{equation}
\hat{Z}(y)\hat{a}_{12}(y) = R(y, \bar{C}), \label{2.10}
\end{equation}
where
\begin{equation}
R(y, \bar{C}) = -(1 + \hat{\lambda})Re\sigma_{m}(\hat{L}(y,\bar{C}) + J^{+}) + \hat{D}(\frac{1}{2} - y) + \bar{C}. \label{2.11}
\end{equation}
Here $\bar{C}(= \hat{a}_{12}(\frac{1}{2}))$ is the constant to be found later on.

\begin{remark}
Note that in \eqref{2.10} we can assume that $\hat{D} = 1$ (due to the choice $u_{H} = \frac{l\widehat{\Delta p}}{h\eta_{0}^{*}}$ of the characteristic parameter $u_{H}$).
\end{remark}

Later on, equations \eqref{2.4} -- \eqref{2.11} enable us to formulate an iterative algorithm for finding $\hat{Z}(y,\bar{C})$ (to be exact, $\hat{Q}(y,\bar{C})$: $\hat{Z}(y,\bar{C}) = \hat{Q}(y,\bar{C}) + 1 + \bar{\theta}(\frac{1}{2} - y)$), $\hat{L}(y,\bar{C})$, $\hat{a}_{22}(y,\bar{C})$ and the constant $\bar{C}$ itself. From formulas \eqref{2.3} -- \eqref{2.6}, \eqref{2.9}, \eqref{2.10} we can find the rest unknown values $\hat{P}$, $\hat{u}$, $\hat{a}_{11}$, $\hat{a}_{12}$ and $\hat{M}$. This algorithm as well as  the results of computations for the values $\hat{u}$, $\hat{Z}$ and  $\hat{L}$ are described in \cite{BS}.

Some solutions or, more precisely, the components of solutions $\hat{u}$, $\hat{Z}$ and  $\hat{L}$ are shown in Fig. 2--5. Moreover,  in the first (main) case $\hat{A} = 1$, $\hat{\lambda} = 1$, $\sigma_{m} = 1$, $Re = 1$, $W = 1$, $\beta = 0.5$, $A_{r} = 1$, $A_{m} = 1$, $\bar{\theta} = 1$, $b_{m} = 1$, $\bar{E}_{A} = 1$, $J^{+} = 2$, $J^{-} = 1$, and in the second, third and fourth cases one of the parameters $A$, $J^{+}$ and $\theta$ is changed respectively, and other parameters stay unchanged.

\begin{figure}[ph]
\centerline{\psfig{file=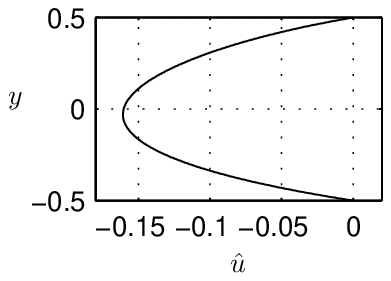,width=1in}\psfig{file=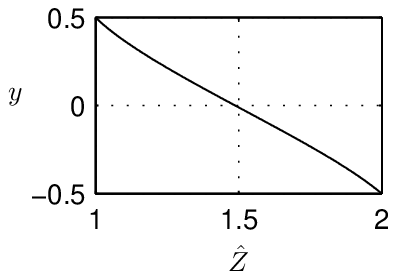,width=1in}\psfig{file=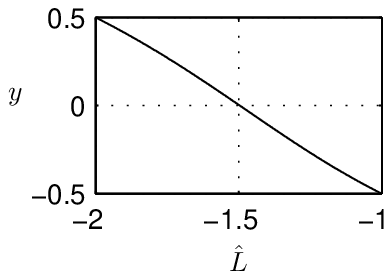,width=1in}} 
\vspace*{8pt}
\caption{Solutions for base values of parameters.}
\end{figure}

\begin{figure}[ph]
\centerline{\psfig{file=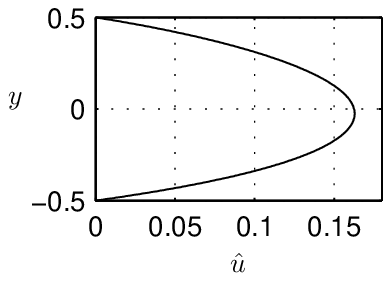,width=1in}\psfig{file=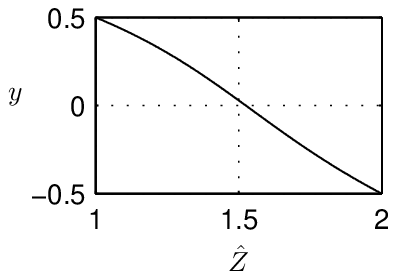,width=1in}\psfig{file=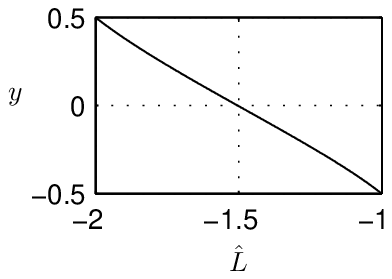,width=1in}} 
\vspace*{8pt}
\caption{Solution for $\hat{A}=3$.}
\end{figure}

\begin{figure}[ph]
\centerline{\psfig{file=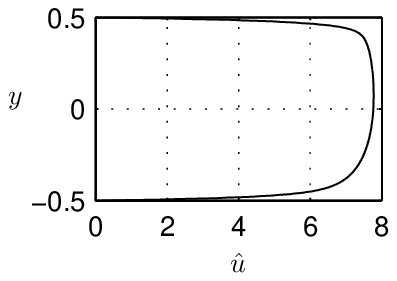,width=1in}\psfig{file=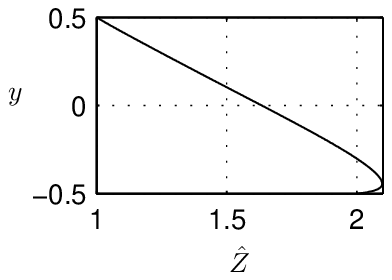,width=1in}\psfig{file=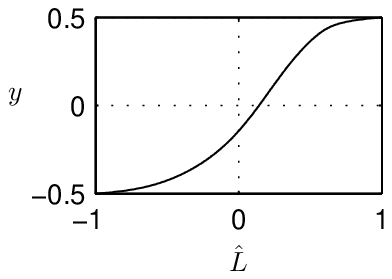,width=1in}} 
\vspace*{8pt}
\caption{Solution for $J^{+}=-1$.}
\end{figure}

\begin{figure}[ph]
\centerline{\psfig{file=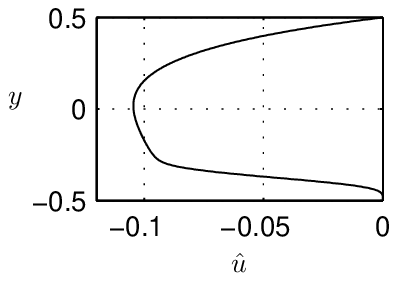,width=1in}\psfig{file=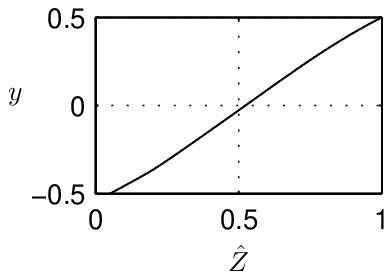,width=1in}\psfig{file=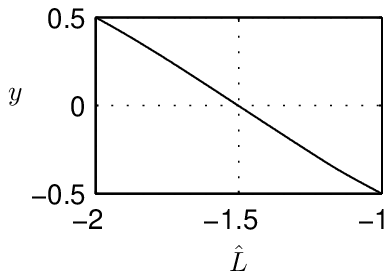,width=1in}} 
\vspace*{8pt}
\caption{Solution for $\bar{\theta}=-0.95$.}
\end{figure}


Let us point out some important qualitative features of the behavior of stationary regimes of the flow of a polymeric fluid. In \cite{6} the influence of the parameter $\bar{E}_{A}$ connected with the activation energy $E_{A}$  ($\bar{E}_{A} = \frac{E_{A}}{T_{0}}$) on the form of the velocity profile was found: the velocity profile looses symmetry natural for the parabolic velocity profile for the Poiseuille flow of a viscous fluid \cite{3}. In our case, solutions of model \eqref{1.1}--\eqref{1.12} have even more interesting features. For instance, due to the influence of magnetic field,  in the main case (see Fig. 2) the velocity profile is stretched in the direction opposite to that of pressure forces. In Fig. 3 the pressure drop is relatively big (according its absolute value) and, as a result, the velocity profile is turned to the right. In Fig. 4 the absolute value of the pressure is small enough, but due to the change of current's direction on the top electrode the velocity profile is again turned to the right.At last, strong cooling of the bottom boundary results in the decay of the fluid velocity at the bottom of the channel (see Fig. 5).

\begin{remark}
In \cite{BS}  Couette-type flows were found numerically whereas in this work as the basic  solution we take a Poiseuille-type flow. i.e., a solution of problem \eqref{1.1}--\eqref{1.12}.
\end{remark}

The most important question appearing for the stationary solution of problem \eqref{1.1} -- \eqref{1.12} (Poiseuille-type flow) is its physical realizability and, in particular, its Lyapunov's stability.

\section{Linearized model of the flow of a polymeric fluid in a plane channel. Formulation of main results}

After linearizing model \eqref{1.1} -- \eqref{1.12} with respect to the Poiseuille-type solution described in the previous section and taking into account additional conditions \eqref{1.2} we get the following initial boundary value problem:
\begin{equation}
\vec{V}_{t} + B\vec{V}_{x} +C\vec{V}_{y} + R\vec{V} + \vec{\Gamma} = 0, \label{3.1}
\end{equation}
\begin{multline}
\Delta\Omega = \hat{Z}[(\alpha_{11})_{xx} - (\alpha_{22})_{xx}] + \frac{2\hat{g}}{\hat{A}_{2}Re}Z_{xx} + 2(\hat{Z}\alpha_{12} + \hat{\alpha}_{12}Z)_{xy} - 2\hat{u}'v_{x} +\\
+2 \sigma_{m}\hat{L}'\frac{\partial M}{\partial x} + GrZ_{y}, \label{3.2}
\end{multline}
\begin{multline}
Z_{t} + \hat{u}Z_{x} + \hat{Z}'v = \frac{1}{Pr}\Delta_{x,y}Z + \frac{Ar}{Pr}\hat{u}'\hat{a}_{12}Z + \frac{Ar}{Pr}\hat{Z}(\hat{a}_{11}u_{x} +\\
+ \hat{a}_{12}v_{x} + \hat{a}_{12}u_{y} + \hat{a}_{22}v_{y} + \hat{u}'a_{12}) + \frac{Am}{Pr}\sigma_{m}(\hat{L}^{2}u_{x} +\\
+ \hat{L}(1 + \hat{\lambda})v_{x} + \hat{L}(1 + \hat{\lambda})u_{y} + \hat{u}'((1 + \hat{\lambda})L + \hat{L}M) +\\
+ (1 + \hat{\lambda})^{2}v_{y}), \label{3.3}
\end{multline}
\begin{equation}
L_{t} + \hat{u}L_{x} + v\hat{L}' - \hat{L}u_{x} - (1 + \hat{\lambda})u_{y} - \hat{u}'M - b_{m}\Delta_{x,y}L = 0, \label{3.4}
\end{equation}
\begin{equation}
M_{t} + \hat{u}M_{x} - \hat{L}v_{x} - (1 + \hat{\lambda})v_{y} - b_{m}\Delta_{x,y}M = 0, \quad -\frac{1}{2} < y < \frac{1}{2}, \label{3.5}
\end{equation}
где
\[
V = (u, v, a_{11}, a_{12}, a_{22})^{T}, \quad B = \hat{U}I_{5} + B_{0},
\]
\[
B_{0} = \begin{pmatrix}
0 & 0 & -\hat{Z} & 0 & Z\\
0 & 0 & 0 & -\hat{Z} & 0\\
-2\hat{\alpha}_{1} & 0 & 0 & 0 & 0\\
0 & -\hat{\alpha}_{1} & 0 & 0 & 0\\
0 & -2\hat{\alpha}_{12} & 0 & 0 & 0
\end{pmatrix}, \quad
C = \begin{pmatrix}
0 & 0 & 0 & -\hat{Z} & 0\\
0 & 0 & 0 & 0 & 0\\
-2\hat{\alpha}_{12} & 0 & 0 & 0 & 0\\
-\hat{\alpha}_{2} & 0 & 0 & 0 & 0\\
0 & -2\hat{\alpha}_{2} & 0 & 0 & 0
\end{pmatrix},
\]
\[
R = \begin{pmatrix}
0 & \hat{u}' & 0 & -\hat{Z}' & 0\\
0 & 0 & 0 & 0 & 0\\
0 & \hat{\alpha}_{11}' & R_{33} & R_{34} & R_{35}\\
0 & \hat{\alpha}_{12}' & R_{43} & R_{44} & R_{45}\\
0 & \hat{\alpha}_{22}' & R_{53} & R_{54} & R_{55}
\end{pmatrix}, \quad
\vec{\Gamma} = \begin{pmatrix}
\Gamma_{1}\\
\Gamma_{2}\\
r_{11}Z\\
r_{12}Z\\
0
\end{pmatrix},
\]
$\Omega = P - \hat{Z}\alpha_{22} - Z\hat{\alpha}_{22}$, $\hat{g} = \hat{a}_{12}$, $\hat{\alpha}_{ij} = \frac{\hat{a}_{ij}}{Re}$,\\
$\Gamma_{1} = \Omega_{x} + Z_{x}(\hat{\alpha}_{22} - \hat{\alpha}_{11}) - \hat{\alpha}_{12}Z_{y} - \hat{\alpha}_{12}'Z + \sigma_{m}(1 + \hat{\lambda})\omega_{m} - \sigma_{m}\hat{L}M$,\\
$\Gamma_{2} = \Omega_{y} - \hat{\alpha}_{12}Z_{x} - GrZ - \sigma_{m}\hat{L}\omega_{m} + \sigma_{m}\hat{L}'L$,\\
$R_{33} = \hat{\chi}_{0}^{*}(\hat{K}_{\hat{I}} + \hat{a}_{11}(\frac{\bar{k}}{3} + 2\beta))$, $R_{34} = -2\hat{u}' + 2\beta\hat{a}_{12}\chi_{0}^{*}$,\\
$R_{35} = \frac{\bar{k}}{3}\hat{a}_{11}\hat{\chi}_{0}^{*}$, $\hat{K}_{I} = \frac{1}{W} + \frac{\bar{k}}{3}(\hat{a}_{11} + \hat{a}_{22})$, $\hat{\chi}_{0}^{*} = \frac{1}{\bar{\tau}_{0}(\hat{Z})}$,\\
$r_{11} = (\hat{\chi}_{0}^{*})'\frac{2\hat{\alpha}_{12}^{2}\tilde{K}_{\hat{I}}}{\hat{\alpha}_{2}}$, $\hat{\tilde{K}}_{\hat{I}} = \hat{K}_{\hat{I}} + \beta(\hat{a}_{11} + \hat{a}_{22})$,\\
$R_{43} = \hat{a}_{12}\hat{\chi}_{0}^{*}(\frac{\bar{k}}{3} + \beta)$, $R_{45} = -\hat{u}' + \hat{a}_{12}\hat{\chi}_{0}^{*}(\frac{\bar{k}}{3} + \beta)$, $R_{44} = \hat{\chi}_{0}^{*}\hat{\tilde{K}}_{\hat{I}}$,\\
$r_{12} = (\hat{\chi}_{0}^{*})'\hat{\alpha}_{12}\hat{\tilde{K}}_{\hat{I}}$,\\
$R_{53} = \hat{\chi}_{0}^{*}\hat{a}_{22}\frac{\bar{k}}{3}$, $R_{54} = 2\beta\hat{a}_{12}\hat{\chi}_{0}^{*}$, $R_{55} = \hat{\chi}_{0}^{*}(\hat{K}_{\hat{I}} + \hat{a}_{22}(\frac{\bar{k}}{3} + 2\beta))$,\\
$\hat{\alpha}_{i} = \frac{\hat{a}_{ii} + W^{-1}}{Re} = \hat{\alpha}_{ii} + \kappa^{2}$, $\kappa^{2} = \frac{1}{WRe}$.

Adding boundary conditions and initial data to system \eqref{3.1} - \eqref{3.5} we get:
\begin{equation}
y = -\frac{1}{2}: \quad u = v = Z = L = M = 0, \quad \Omega_{y} = (1 + \bar{\theta})(\alpha_{12})_{x}; \label{3.6}
\end{equation}
\begin{equation}
y = \frac{1}{2}: \quad u = v = Z = L = M = 0, \quad \Omega_{y} = (\alpha_{12})_{x}. \label{3.7}
\end{equation}
Moreover, the initial data must satisfy equation \eqref{3.2} and the additional condition \eqref{1.2}.

We will seek partial solutions of the above problem in the special form
\begin{multline}
\begin{aligned}
V &= \exp(\lambda t + i\omega x)V(y),\\
(\Omega, Z, L, M) &= \exp(\lambda t + i\omega x)(\Omega^{*}(y), Z^{*}(y), L^{*}(y), M^{*}(y)),\\
\lambda &= \eta + i\xi, \quad \xi, \omega \in R^{1}.
\end{aligned} \label{3.8}
\end{multline}
(here and below hats are dropped).

Let us formulate our  main result.

\begin{theorem}
If problem \eqref{3.1} -- \eqref{3.7} has solutions of the form \eqref{3.8} (the parameter $\omega$ is fixed), then we have the following asymptotic representation for $\lambda$:
\begin{multline}
\lambda_{k} = \left[\int_{-\frac{1}{2}}^{\frac{1}{2}}\frac{1}{\sqrt{\hat{Z}\hat{\alpha}_{2}}}d\xi\right]^{-1}
\left(\int_{-\frac{1}{2}}^{\frac{1}{2}}\frac{1}{\sqrt{\hat{Z}\hat{\alpha}_{2}}} (i\omega\hat{u} + \frac{R_{43}\hat{\alpha}_{12}}{\hat{\alpha}_{2}} + \frac{R_{44}}{2})d\xi + k\pi i\right) + O(\frac{1}{k}), \\
k \to \infty , \label{3.9}
\end{multline}
\end{theorem}
where $O (\cdot )$ is the usual big $O$ notation.

From representation \eqref{3.9} we find the necessary condition for the asymptotic Lyapunov's stability of the Poiseuille-type flows found in Sect. 2.

\begin{theorem}

The fulfilment of the inequality 
\begin{equation}
\int_{-\frac{1}{2}}^{\frac{1}{2}}\frac{\hat{\chi}_{0}^{*}}{\sqrt{\hat{Z}\hat{\alpha}_{2}}}(\hat{a}_{11}(\frac{\bar{k}}{3} + \beta) + \frac{1}{2W})d\xi < 0 \label{3.10}
\end{equation}
is necessary for the asymptotic Lyapunov's stability of the Poiseuille-type flow.
\end{theorem}

\section{Proof of Theorems 1 and 2}

Following  \cite{15}, \cite{16} and using representation \eqref{3.8} of the solution, let us make the following change of unknowns in system \eqref{3.1}:
\begin{equation}
V = TY = T\begin{pmatrix} y_{1}\\ y_{2}\\ y_{3}\\ y_{4}\\ y_{5} \end{pmatrix}, \label{4.1}
\end{equation}
where
\begin{equation}
T = \begin{pmatrix}
0 & 0 & 0 & \sqrt{\frac{\hat{Z}}{\hat{\alpha}_{2}}} & \sqrt{\frac{\hat{Z}}{\hat{\alpha}_{2}}}\\
0 & 0 & \frac{1}{2\hat{\alpha}_{2}} & 0 & 0\\
1 & 0 & 0 & 2\frac{\hat{\alpha}_{12}}{\hat{\alpha}_{2}} & \frac{\hat{\alpha}_{12}}{\hat{\alpha}_{2}}\\
0 & 0 & 0 & 1 & 1\\
0 & 1 & 0 & 0 & 0
\end{pmatrix}. \label{4.2}
\end{equation}
Then, \eqref{3.1} takes the following form:
\begin{equation}
\{\lambda I_{5} + K\frac{d}{dy} + (i\omega L + T^{-1}(RT + CT'))\}Y + \tilde{\Gamma} = 0. \label{4.3}
\end{equation}
Here
\begin{equation}
K = \begin{pmatrix}
0 & 0 & 0 & 0 & 0\\
0 & 0 & 1 & 0 & 0\\
0 & 0 & 0 & 0 &0\\
0 & 0 & 0 & \sqrt{\hat{Z}\hat{\alpha}_{2}} & 0\\
0 & 0 & 0 & 0 & -\sqrt{\hat{Z}\hat{\alpha}_{2}}
\end{pmatrix}, \label{4.4}
\end{equation}
\[
L = \begin{pmatrix}
\hat{u} & 0 & \frac{\hat{\alpha}_{12}\hat{\alpha}_{1}}{\hat{\alpha}_{2}^{2}} & 2\hat{\alpha}_{1}\sqrt{\frac{\hat{Z}}{\hat{\alpha}_{2}}} & - 2\hat{\alpha}_{1}\sqrt{\frac{\hat{Z}}{\hat{\alpha}_{2}}}\\
0 & \hat{u} & \frac{\hat{\alpha}_{12}}{\hat{\alpha}_{2}} & 0 & 0\\
0 & 0 & \hat{u}' & 2\hat{\alpha}_{2}\hat{Z} & 2\hat{\alpha}_{2}\hat{Z}\\
\frac{1}{2}\sqrt{\hat{Z}\hat{\alpha}_{1}} & -\frac{1}{2}\sqrt{\hat{Z}\hat{\alpha}_{2}} & \frac{1}{4}\frac{\hat{\alpha}_{1}}{\hat{\alpha}_{2}} & \hat{u} + \hat{\alpha}_{12}\sqrt{\frac{\hat{Z}}{\hat{\alpha}_{2}}} & \hat{\alpha}_{12}\sqrt{\frac{\hat{Z}}{\hat{\alpha}_{2}}}\\
-\frac{1}{2}\sqrt{\hat{Z}\hat{\alpha}_{1}} & \frac{1}{2}\sqrt{\hat{Z}\hat{\alpha}_{2}} & \frac{1}{4}\frac{\hat{\alpha}_{1}}{\hat{\alpha}_{2}} & - \hat{\alpha}_{12}\sqrt{\frac{\hat{Z}}{\hat{\alpha}_{2}}} & \hat{u} - \hat{\alpha}_{12}\sqrt{\frac{\hat{Z}}{\hat{\alpha}_{2}}}
\end{pmatrix},
\]
$T^{-1}RT + T^{-1}CT' = S$, where
\begin{multline}
\begin{aligned}
s_{11} &= R_{33} - \frac{2\hat{\alpha}_{12}R_{43}}{\hat{\alpha}_{2}}, \, s_{12} = R_{35} - \frac{2\hat{\alpha}_{12}R_{45}}{\hat{\alpha}_{2}}, \, s_{13} = -\frac{\hat{\alpha}_{11}'}{2\hat{\alpha}_{2}} + \frac{\hat{\alpha}_{12}\hat{\alpha}_{12}'}{\hat{\alpha}_{2}^{2}},\\
s_{14} &= \frac{2\hat{\alpha}_{12}R_{33}}{\hat{\alpha}_{2}} + R_{34} - \frac{4\hat{\alpha}_{12}^{2}R_{43}}{\hat{\alpha}^{2}_{2}} - \frac{2\hat{\alpha}_{12}R_{44}}{\hat{\alpha}_{2}},\\
s_{15} &= \frac{2\hat{\alpha}_{12}R_{33}}{\hat{\alpha}_{2}} + R_{34} - \frac{4\hat{\alpha}_{12}^{2}R_{43}}{\hat{\alpha}^{2}_{2}} - \frac{2\hat{\alpha}_{12}R_{44}}{\hat{\alpha}_{2}};\\
s_{21} &= R_{53}, \, s_{22} = R_{55}, \, s_{23} = -\frac{\hat{\alpha}_{22}'}{2\hat{\alpha}_{2}} + \hat{\alpha}_{2}(\frac{1}{\hat{\alpha}_{2}})', \, s_{24} = R_{54} + \frac{2R_{53}\hat{\alpha}_{12}}{\hat{\alpha}_{2}}, \\
s_{25} &= R_{54} + \frac{2R_{53}\hat{\alpha}_{12}}{\hat{\alpha}_{2}};\\
s_{31} &= s_{32} = s_{33} = s_{34} = s_{35} = 0;\\
s_{41} &= \frac{R_{43}}{2}, \, s_{42} = \frac{R_{45}}{2}, \, s_{43} = \frac{1}{4}\hat{u}'\sqrt{\frac{1}{\hat{Z}\hat{\alpha}_{2}}} - \frac{1}{4}\frac{\hat{\alpha}_{12}'}{\hat{\alpha}_{2}}, \, s_{44} = \frac{1}{2}\hat{Z}'\sqrt{\frac{\hat{\alpha}_{2}}{\hat{Z}}} +\\
 &+ \frac{R_{43}\hat{\alpha}_{12}}{\hat{\alpha}_{2}} + \frac{R_{44}}{2} +\\
&+ \frac{\hat{\alpha}_{2}}{2}\left(\sqrt{\frac{\hat{Z}}{\hat{\alpha}_{2}}}\right)', \,
s_{45} = \frac{1}{2}\hat{Z}'\sqrt{\frac{\hat{\alpha}_{2}}{\hat{Z}}} + \frac{R_{43}\hat{\alpha}_{12}}{\hat{\alpha}_{2}} + \frac{R_{44}}{2} - \frac{\hat{\alpha}_{2}}{2}\left(\sqrt{\frac{\hat{Z}}{\hat{\alpha}_{2}}}\right)';\\
s_{51} &= \frac{R_{43}}{2}, \, s_{52} = \frac{R_{45}}{2}, \, s_{53} = -\frac{1}{4}\hat{u}'\sqrt{\frac{1}{\hat{Z}\hat{\alpha}_{2}}} - \frac{1}{4}\frac{\hat{\alpha}_{12}'}{\hat{\alpha}_{2}}, \, s_{54} = -\frac{1}{2}\hat{Z}'\sqrt{\frac{\hat{\alpha}_{2}}{\hat{Z}}} +\\
&+ \frac{R_{43}\hat{\alpha}_{12}}{\hat{\alpha}_{2}} + \frac{R_{44}}{2} +\\
&+ \frac{\hat{\alpha}_{2}}{2}\left(\sqrt{\frac{\hat{Z}}{\hat{\alpha}_{2}}}\right)', \,
s_{55} = -\frac{1}{2}\hat{Z}'\sqrt{\frac{\hat{\alpha}_{2}}{\hat{Z}}} + \frac{R_{43}\hat{\alpha}_{12}}{\hat{\alpha}_{2}} + \frac{R_{44}}{2} - \frac{\hat{\alpha}_{2}}{2}\left(\sqrt{\frac{\hat{Z}}{\hat{\alpha}_{2}}}\right)'.
\end{aligned} \label{4.5}
\end{multline}
The components of the vector $\tilde{\Gamma}$ read:
\begin{multline*}
\begin{aligned}
\tilde{\Gamma}_{1} &= (r_{11} - \frac{\hat{\alpha}_{12}}{\hat{\alpha}_{2}}r_{12})Z^{*};\\
\tilde{\Gamma}_{2} &= 0; \, \tilde{\Gamma}_{3} = -2\hat{\alpha}_{2}(\Omega^{*}_{y} - (\hat{\alpha}_{12}i\omega + Gr)Z^{*} - \sigma_{m}i\omega\hat{L}M^{*} + \sigma_{m}\hat{L}L_{y}^{*} + \sigma_{m}\hat{L}'L^{*});\\
\tilde{\Gamma}_{4} &= -\frac{1}{2}\sqrt{\frac{\hat{\alpha}_{2}}{\hat{Z}}}i\omega\Omega^{*} + Z^{*}\{\frac{1}{2}\sqrt{\frac{\hat{\alpha}_{2}}{\hat{Z}}}((\hat{\alpha}_{11} - \hat{\alpha}_{22})i\omega + \hat{\alpha}_{12}') + \frac{1}{2}r_{12}\} + \frac{1}{2}\hat{\alpha}_{12}\sqrt{\frac{\hat{\alpha}_{2}}{\hat{Z}}}Z_{y}^{*} -\\
&- \frac{1}{2}\sqrt{\frac{\hat{\alpha}_{2}}{\hat{Z}}}\sigma_{m}((1 + \hat{\lambda})i\omega - \hat{L}')M^{*} + \frac{\sigma_{m}}{2}\sqrt{\frac{\hat{\alpha}_{2}}{\hat{Z}}}(1 + \hat{\lambda})L_{y}^{*};\\
\tilde{\Gamma}_{5} &= -\frac{1}{2}\sqrt{\frac{\hat{\alpha}_{2}}{\hat{Z}}}i\omega\Omega^{*} + Z^{*}\{-\frac{1}{2}\sqrt{\frac{\hat{\alpha}_{2}}{\hat{Z}}}((\hat{\alpha}_{11} - \hat{\alpha}_{22})i\omega + \hat{\alpha}_{12}') + \frac{1}{2}r_{12}\} - \frac{1}{2}\hat{\alpha}_{12}\sqrt{\frac{\hat{\alpha}_{2}}{\hat{Z}}}Z_{y}^{*} +\\
&+ \frac{1}{2}\sqrt{\frac{\hat{\alpha}_{2}}{\hat{Z}}}\sigma_{m}((1 + \hat{\lambda})i\omega - \hat{L}')M^{*} - \frac{\sigma_{m}}{2}\sqrt{\frac{\hat{\alpha}_{2}}{\hat{Z}}}(1 + \hat{\lambda})L_{y}^{*}.
\end{aligned}
\end{multline*}
Equations \eqref{3.2}--\eqref{3.5} take the new form
\begin{multline}
(\Omega^{*})'' - \omega^{2}\Omega^{*} = -\omega^{2}\hat{Z}y_{1} + \omega^{2}\hat{Z}y_{2} + \frac{i\omega\hat{u}'}{\hat{\alpha}_{2}}y_{3} + y_{4}(-\omega^{2}\hat{Z}\frac{2\hat{\alpha}_{12}}{\hat{\alpha}_{2}} + 2i\omega\hat{Z}') +\\
+ y_{5}(-\omega^{2}\hat{Z}\frac{2\hat{\alpha}_{12}}{\hat{\alpha}_{2}} + 2i\omega\hat{Z}') + 2i\omega\hat{Z}y_{4}' + 2i\omega\hat{Z}y_{5}' +\\
+ \sigma_{m}2i\omega\hat{L}M^{*} + (2i\omega\hat{\alpha}_{12} + Gr)(Z^{*})', \label{4.6}
\end{multline}
\begin{multline}
\frac{1}{Pr}(Z^{*})'' + (-\frac{\omega^{2}}{Pr} - \lambda - i\omega\hat{u} + \frac{A_{r}}{Pr}\hat{u}'\hat{a}_{12})Z^{*} - (\hat{Z}' -i\omega\frac{\hat{a}_{12}A_{r}\hat{Z}}{Pr} - \frac{A_{m}}{Pr}\sigma_{m}\hat{L}(1+ \hat{\lambda})i\omega)\times\\
\times(-\frac{1}{2\alpha_{2}}y_{3}) + (i\omega\hat{a}_{11}\frac{A_{r}\hat{Z}}{Pr} + \frac{A_{m}}{Pr}\sigma_{m}i\omega\hat{L}^{2})(-\sqrt{\frac{\hat{Z}}{\hat{\alpha}_{2}}}y_{4} + \sqrt{\frac{\hat{Z}}{\hat{\alpha}_{2}}}y_{5}) +\\
+ (\frac{\hat{a}_{12}A_{r}\hat{Z}}{Pr} + \frac{A_{m}}{Pr}\sigma_{m}\hat{L}(1 + \hat{\lambda}))(-\left(\sqrt{\frac{\hat{Z}}{\hat{\alpha}_{2}}}\right)'y_{4} - \sqrt{\frac{\hat{Z}}{\hat{\alpha}_{2}}}y_{4}' + \left(\sqrt{\frac{\hat{Z}}{\hat{\alpha}_{2}}}\right)'y_{5} +\\
+ \sqrt{\frac{\hat{Z}}{\hat{\alpha}_{2}}}y_{5}') + (\frac{\hat{a}_{22}A_{r}\hat{Z}}{Pr} + \frac{A_{m}\sigma_{m}}{Pr}(1 + \hat{\lambda})^{2})(\frac{\hat{\alpha}_{2}'}{2\hat{\alpha}_{2}^{2}}y_{3} - \frac{1}{2\hat{\alpha}_{2}}y_{3}') +\\
 + \frac{\hat{u}'A_{r}\hat{Z}}{Pr}(y_{4} + y_{5}) + \frac{A_{m}}{Pr}\sigma_{m}\hat{u}'(1 + \hat{\lambda})L^{*} + \frac{A_{m}}{Pr}\sigma_{m}\hat{u}'\hat{L}M^{*} = 0, \label{4.7}
\end{multline}
\begin{multline}
-b_{m}(L^{*})'' + b_{m}(\omega^{2} + \lambda + i\omega\hat{u})L^{*} - y_{4}(-i\omega\hat{L}\sqrt{\frac{\hat{Z}}{\hat{\alpha}_{2}}} - (1 + \hat{\lambda})\left(\sqrt{\frac{\hat{Z}}{\hat{\alpha}_{2}}}\right)') -\\
- y_{5}\bigg(\left(\sqrt{\frac{\hat{Z}}{\hat{\alpha}_{2}}}\right)'(1 + \hat{\lambda}) + \sqrt{\frac{\hat{Z}}{\hat{\alpha}_{2}}}i\omega\hat{L}\bigg) + y_{4}'\sqrt{\frac{\hat{Z}}{\hat{\alpha}_{2}}}(1 + \hat{\lambda}) - y_{5}'\sqrt{\frac{\hat{Z}}{\hat{\alpha}_{2}}}(1 + \hat{\lambda}) +\\
+ \hat{L}'\frac{\hat{\alpha}_{2}'}{2\hat{\alpha}_{2}^{2}}y_{3} - \hat{L}'\frac{1}{2\hat{\alpha}_{2}}y_{3}' - \hat{u}'M^{*} = 0, \label{4.8}
\end{multline}
\begin{equation}
-b_{m}(M^{*})'' + b_{m}(\omega^{2} + \lambda + i\omega\hat{u})M^{*} + y_{3}(\frac{i\omega\hat{L}}{2\hat{\alpha}_{2}} - \frac{(1 + \hat{\lambda})\hat{\alpha}_{2}'}{2\hat{\alpha}_{2}^{2}}) + \frac{1 + \hat{\lambda}}{2\hat{\alpha}_{2}}y_{3}' = 0, \label{4.9}
\end{equation}
as well as the boundary conditions \eqref{3.6}, \eqref{3.7}:
\begin{equation}
y = -\frac{1}{2}: \quad y_{4} = y_{5}, \quad Z^{*} = L^{*} = M^{*} = 0, \, (\Omega^{*})' = (1 + \bar{\theta})i\omega(y_{4} + y_{5}); \label{4.10}
\end{equation}
\begin{equation}
y = \frac{1}{2}: \quad y_{4} = y_{5}, \quad Z^{*} = L^{*} = M^{*} = 0, \, (\Omega^{*})' = i\omega(y_{4} + y_{5}). \label{4.11}
\end{equation}
In addition to the change of unknowns \eqref{4.1} let us also denote
\begin{multline}
\Omega^{*} = y_{6},\, (\Omega^{*})' = y_{7}, \, Z^{*} = y_{8}, \, (Z^{*})' = y_{9}, \, L^{*} = y_{10}, \, (L^{*})' = y_{11}, \, M^{*} = y_{12}, \\
 (M^{*})' = y_{13}. \label{4.12}
\end{multline}
Taking into account the structure of the matrices $K$ and $P$, we can express the first three components $y_{1}$, $y_{2}$ and $y_{3}$ through the rest ones. Indeed,
\begin{multline}
y_{1} = \frac{1}{\lambda + i\omega\hat{u} + R_{33} - \frac{2\hat{\alpha}_{12}R_{43}}{\hat{\alpha}_{2}}}\bigg[(-R_{35} + \frac{2\hat{\alpha}_{12}R_{45}}{\hat{\alpha}_{2}})y_{2} + (-\frac{\hat{\alpha}_{12}\hat{\alpha}_{1}}{\hat{\alpha}_{2}^{2}}i\omega + \frac{\hat{\alpha}_{11}'}{2\hat{\alpha}_{2}} +\\
+ \frac{\hat{\alpha}_{12}\hat{\alpha}_{12}'}{\hat{\alpha}_{2}^{2}})y_{3} + (-2i\omega\alpha_{1}\sqrt{\frac{\hat{Z}}{\hat{\alpha}_{2}}} - 2\frac{\hat{\alpha}_{12}R_{33}}{\hat{\alpha}_{2}} - R_{34} + 4\frac{\hat{\alpha}_{12}^{2}R_{43}}{\hat{\alpha}_{2}^{2}} + \frac{2\hat{\alpha}_{12}R_{44}}{\hat{\alpha}_{2}})y_{4} +\\
+ (2i\omega\alpha_{1}\sqrt{\frac{\hat{Z}}{\hat{\alpha}_{2}}} - 2\frac{\hat{\alpha}_{12}R_{33}}{\hat{\alpha}_{2}} - R_{34} + 4\frac{\hat{\alpha}_{12}^{2}R_{43}}{\hat{\alpha}_{2}^{2}} + \frac{2\hat{\alpha}_{12}R_{44}}{\hat{\alpha}_{2}})y_{5} - \\
- (r_{11} - \frac{2\hat{\alpha}_{12}}{\hat{\alpha}_{2}}r_{12})y_{8}\bigg]. \label{4.13}
\end{multline} 
\begin{multline}
y_{3} = \frac{1}{\lambda + i\omega\hat{u}}\bigg[-2i\omega\hat{Z}\hat{\alpha}_{2}y_{4} - 2i\omega\hat{Z}\hat{\alpha}_{2}y_{5} + 2\hat{\alpha}_{2}y_{7} + 2\hat{\alpha}_{2}(\hat{\alpha}_{12}i\omega + Gr)y_{8} +\\
+ \sigma_{m}\hat{L}'2\hat{\alpha}_{2}y_{10} + \sigma_{m}\hat{L}2\hat{\alpha}_{2}y_{11} - \sigma_{m}i\omega\hat{L}2\hat{\alpha}_{2}y_{12}\bigg]. \label{4.14}
\end{multline}
The incompressibility condition in terms of the new unknown \eqref{4.1} is written as
\[
(-\frac{1}{2\hat{\alpha}_{2}}y_{3})' + (-\sqrt{\frac{\hat{Z}}{\hat{\alpha}_{2}}}y_{4} + \sqrt{\frac{\hat{Z}}{\hat{\alpha}_{2}}}y_{5})i\omega = 0.
\]
Consequently, from the last equality and the second equation of system \eqref{4.3} we get:
\begin{multline}
-\frac{1}{2\hat{\alpha}_{2}}\bigg(-R_{53}y_{1} - (\lambda + i\omega + R_{55})y_{2} -(i\omega\frac{\hat{\alpha}_{12}}{\hat{\alpha}_{2}} - \frac{\hat{\alpha}_{22}'}{2\hat{\alpha}_{2}} + \hat{\alpha}_{2}(\frac{1}{\hat{\alpha}_{2}})')y_{3} -\\
- (R_{54} + \frac{2R_{53}\hat{\alpha}_{12}}{\hat{\alpha}_{2}})y_{4} - (R_{54} + \frac{2R_{53}\hat{\alpha}_{12}}{\hat{\alpha}_{2}})y_{5}\bigg) - (\frac{1}{2\hat{\alpha}_{2}})'\frac{1}{\lambda + i\omega\hat{u}}\times\\
\times\bigg(-2\hat{\alpha}_{2}i\omega\hat{Z}y_{4} - 2\hat{\alpha}_{2}i\omega\hat{Z}y_{5} + 2\hat{\alpha}_{2}y_{7} + 2\hat{\alpha}_{2}(\hat{\alpha}_{12}i\omega + Gr)y_{8} +\\
+ \sigma_{m}\hat{L}'2\hat{\alpha}_{2}y_{10} + \sigma_{m}\hat{L}2\hat{\alpha}_{2}y_{11} - \sigma_{m}i\omega\hat{L}2\hat{\alpha}_{2}y_{12}\bigg) = i\omega\sqrt{\frac{\hat{Z}}{\hat{\alpha}_{2}}}y_{4} - i\omega\sqrt{\frac{\hat{Z}}{\hat{\alpha}_{2}}}y_{5}. \label{4.15}
\end{multline}
Substituting $y_{1}$ from \eqref{4.13} and $y_{3}$ from \eqref{4.14} into formula \eqref{4.15}, we get a representation for $y_{2}$ through $y_{4}$, $y_{5}$, \dots, $y_{13}$:
\begin{multline}
y_{2} = -\frac{R_{54} + 2\frac{R_{53}\hat{\alpha}_{12}}{\hat{\alpha}_{2}} - 2i\omega\hat{\alpha}_{2}\sqrt{\frac{\hat{Z}}{\hat{\alpha}_{2}}}}{\lambda  +i\omega + R_{55}}y_{4} - \frac{R_{54} + 2\frac{R_{53}\hat{\alpha}_{12}}{\hat{\alpha}_{2}} + 2i\omega\hat{\alpha}_{2}\sqrt{\frac{\hat{Z}}{\hat{\alpha}_{2}}}}{\lambda  +i\omega + R_{55}}y_{5} +\\
+ O(\frac{1}{\lambda^{2}}), \label{4.16}
\end{multline}
where $O (\cdot )$ is again the usual big $O$ notation. The components $y_{1}$ and $y_{3}$  can be now found by using formulas \eqref{4.13} and \eqref{4.14}.

We will below study the system for the components $y_{4}$, \dots, $y_{13}$, and we will see that the values with the multiplier $\frac{1}{\lambda}$ or its greater powers do not influence on the main term of the asymptotic representation as $|\lambda| \to \infty$, i.e., the components $y_{1}$, $y_{2}$ and $y_{3}$ can be omitted.

In view of system \eqref{4.3} and equations \eqref{4.6}--\eqref{4.9}, the new system for the components $y_{4}$, $y_{5}$, \dots, $y_{13}$ can be written as
\begin{equation}
PY' = GY, \label{4.17}
\end{equation}
where the matrix $P$ is significantly sparse:
\begin{align*}
p_{44} &= \sqrt{\hat{Z}\hat{\alpha}_{2}}, \, p_{55} = -p_{44}, \, p_{66} = p_{77} = p_{88} = p_{10,10} = p_{12, 12} = 1,\\
p_{74} &= -2i\omega\hat{Z}, \, p_{75} = -p_{74}, \, p_{93} = -\frac{1}{2\hat{\alpha}_{2}}\left(\frac{\hat{a}_{22}A_{r}\hat{Z}}{Pr} + \frac{A_{m}\sigma_{m}}{Pr}(1+ \hat{\lambda})^{2}\right),\\
p_{94} &= -\sqrt{\frac{\hat{Z}}{\hat{\alpha}_{2}}}\left(\frac{\hat{a}_{22}A_{r}\hat{Z}}{Pr} + \frac{A_{m}\sigma_{m}}{Pr}(1+ \hat{\lambda})\right), \, p_{95} = -p_{94}, \, p_{99} = \frac{1}{Pr},\\
p_{11, 3} &= -\hat{L}'\frac{1}{2\hat{\alpha}_{2}}, \, p_{11, 4} = \sqrt{\frac{\hat{\alpha}_{2}}{\hat{Z}}}(1 + \hat{\lambda}), \, p_{11, 5} = -p_{11, 4}, \, p_{11, 11} = -b_{m},\\
p_{13, 3} &= \frac{1 + \hat{\lambda}}{2\hat{\alpha}_{2}}, \, p_{13, 13} = -b_{m},
\end{align*}
other elements of $P$ equal zero. The matrix $G$ is rather cumbersome and so we  write down only its elements with the parameter $\lambda$:
\begin{align*}
g_{44} &= -\lambda - i\omega\hat{u} - i\omega\hat{\alpha}_{12}\sqrt{\frac{\hat{Z}}{\hat{\alpha}_{2}}} + \frac{1}{2}\hat{Z}'\sqrt{\frac{\hat{\alpha}_{2}}{\hat{Z}}} - \frac{R_{43}\hat{\alpha}_{12}}{\hat{\alpha}_{12}} - \frac{R_{44}}{2} - \frac{\hat{\alpha}_{2}}{2}\left(\sqrt{\frac{\hat{Z}}{\hat{\alpha}_{2}}}\right)',\\
g_{55} &= -\lambda - i\omega\hat{u} - i\omega\hat{\alpha}_{12}\sqrt{\frac{\hat{Z}}{\hat{\alpha}_{2}}} - \frac{1}{2}\hat{Z}'\sqrt{\frac{\hat{\alpha}_{2}}{\hat{Z}}} - \frac{R_{43}\hat{\alpha}_{12}}{\hat{\alpha}_{12}} - \frac{R_{44}}{2} + \frac{\hat{\alpha}_{2}}{2}\left(\sqrt{\frac{\hat{Z}}{\hat{\alpha}_{2}}}\right)',\\
g_{98} &= \lambda + i\omega\hat{u} - \frac{A_{r}}{Pr}\hat{u}'\hat{a}_{12} + \frac{\omega^{2}}{Pr},\\
g_{11, 10} &= -b_{m}(\omega^{2} + \lambda + i\omega\hat{u}),\\
g_{13, 12} &= -b_{m}(\omega^{2} + \lambda + i\omega\hat{u}).
\end{align*}
In view of our notation \eqref{4.12}, the boundary conditions \eqref{4.10} and \eqref{4.11}  become
\begin{equation}
y = -\frac{1}{2}: \quad y_{4} = y_{5}, \quad y_{8} = y_{11} = y_{12} = 0, \, y_{7} = (1 + \bar{\theta})i\omega(y_{4} + y_{5}); \label{4.18}
\end{equation}
\begin{equation}
y = \frac{1}{2}: \quad y_{4} = y_{5}, \quad y_{8} = y_{11} = y_{12} = 0, \, y_{7} = i\omega(y_{4} + y_{5}). \label{4.19}
\end{equation}
Let us transform the matrix $P$ to the identity matrix $I_{10}$. In that case the matrix $G$ is transformed to the matrix $\tilde{G}$:
\begin{multline}
\begin{aligned}
\tilde{g}_{4, j} &= \frac{1}{\sqrt{\hat{Z}\hat{\alpha}_{2}}}g_{4,j},\\
\tilde{g}_{5,j} &= -\frac{1}{\sqrt{\hat{Z}\hat{\alpha}_{2}}}g_{5,j},\\
\tilde{g}_{7,j} &= g_{7,j} + 2i\omega(\tilde{g}_{4,j} + g_{5,j}),\\
\tilde{g}_{13,j} &= -\frac{1}{b_{m}}g_{13,j},\\
\tilde{g}_{11,j} &= -\frac{1}{b_{m}}(g_{11,j} - \sqrt{\frac{\hat{Z}}{\hat{\alpha}_{2}}}(1 + \hat{\lambda})(\tilde{g}_{4,j} + \tilde{g}_{5,j})),\\ 
\tilde{g}_{9,j} &= Pr(g_{9,j} + (\tilde{g}_{4,j} - \tilde{g}_{5,j})\sqrt{\frac{\hat{Z}}{\hat{\alpha}_{2}}}(\frac{\hat{a}_{12}A_{r}\hat{Z}}{Pr} + \frac{A_{m}\sigma_{m}}{Pr}\hat{L}(1 + \hat{\lambda}))).
\end{aligned} \label{4.20}
\end{multline}
It is convenient to rewrite system \eqref{4.17} by renumbering components of the vector $Y$:
\begin{equation}
I_{10}\tilde{Y}' = Q\tilde{Y}, \quad \tilde{Y} = \begin{pmatrix} \tilde{y}_{1}\\ \vdots\\ \tilde{y}_{10}\end{pmatrix} = \begin{pmatrix} y_{4}\\ \vdots\\ y_{13}\end{pmatrix}. \label{4.21}
\end{equation}
The coefficient matrix by $\lambda$ in the right-hand side of system \eqref{4.21} is as follows: 
\begin{equation}
W = \begin{pmatrix}
-\frac{1}{\sqrt{\hat{Z}\hat{\alpha}_{2}}} & 0 & 0 & 0 & 0 & 0 & 0 & 0 & 0 & 0\\
0 & \frac{1}{\sqrt{\hat{Z}\hat{\alpha}_{2}}} & 0 & 0 & 0 & 0 & 0 & 0 & 0 & 0\\
0 & 0 & 0 & 0 & 0 & 0 & 0 & 0 & 0 & 0\\
-2i\omega\sqrt{\frac{\hat{Z}}{\hat{\alpha}_{2}}} & 2i\omega\sqrt{\frac{\hat{Z}}{\hat{\alpha}_{2}}} & 0 & 0 & 0 & 0 & 0 & 0 & 0 & 0\\
0 & 0 & 0 & 0 & 0 & 0 & 0 & 0 & 0 & 0\\
-Pr\alpha_{0} & -Pr\alpha_{0} & 0 & 0 & Pr & 0 & 0 & 0 & 0 & 0\\
0 & 0 & 0 & 0 & 0 & 0 & 0 & 0 & 0 & 0\\
-\frac{1}{b_{m}}\frac{1 + \hat{\lambda}}{\hat{\alpha}_{2}} & \frac{1}{b_{m}}\frac{1 + \hat{\lambda}}{\hat{\alpha}_{2}} & 0 & 0 & 0 & 0 & 1 & 0 & 0 & 0\\
0 & 0 & 0 & 0 & 0 & 0 & 0 & 0 & 0 & 0\\
0 & 0 & 0 & 0 & 0 & 0 & 0 & 0 & 1 & 0\\
\end{pmatrix}, \label{4.22}
\end{equation}
where $\alpha_{0} = \frac{1}{\hat{\alpha}_{2}}\left(\frac{\hat{a}_{12}A_{r}\hat{Z}}{Pr} + \frac{A_{m}\sigma_{m}}{Pr}\hat{L}(1 + \hat{\lambda})\right)$.

With the help of the new transition matrix $\tilde{T}$ for which
\begin{align*}
t_{11} &= t_{22} = t_{33} = t_{44} = t_{65} = t_{76} = t_{78} = t_{87} = t_{9, 10} = t_{10, 9} = 1,\\
t_{41} &= t_{42} = 2i\omega\hat{Z}, \, t_{56} = \frac{1}{Pr}, \, t_{61} = Pr\alpha_{0}\sqrt{\hat{Z}\hat{\alpha}_{2}}, \, t_{62} = -Pr\alpha_{0}\sqrt{\hat{Z}\hat{\alpha}_{2}},\\
t_{81} &= t_{82} = \frac{1}{b_{m}}(1 + \hat{\lambda})\sqrt{\frac{\hat{Z}}{\hat{\alpha}_{2}}},
\end{align*}
and other elements are zero, the matrix $W$ is transformed to its canonical form
\begin{equation}
\tilde{T}^{-1}W\tilde{T} = \begin{pmatrix}
-\frac{1}{\sqrt{\hat{Z}\hat{\alpha}_{2}}} & 0 & 0 & 0 & 0 & 0 & 0 & 0 & 0 & 0\\
0 & \frac{1}{\sqrt{\hat{Z}\hat{\alpha}_{2}}} & 0 & 0 & 0 & 0 & 0 & 0 & 0 & 0\\
0 & 0 & 0 & 0 & 0 & 0 & 0 & 0 & 0 & 0\\
0 & 0 & 0 & 0 & 0 & 0 & 0 & 0 & 0 & 0\\
0 & 0 & 0 & 0 & 0 & 0 & 0 & 0 & 0 & 0\\
0 & 0 & 0 & 0 & 0 & 0 & 0 & 0 & 0 & 0\\
0 & 0 & 0 & 0 & 0 & 0 & 0 & 1 & 0 & 0\\
0 & 0 & 0 & 0 & 0 & 0 & 0 & 0 & 0 & 0\\
0 & 0 & 0 & 0 & 0 & 0 & 0 & 0 & 0 & 1\\
0 & 0 & 0 & 0 & 0 & 0 & 0 & 0 & 0 & 0\\
\end{pmatrix}. \label{4.23}
\end{equation}
After the change of unknown 
\[
\tilde{Y} = \tilde{T}\hat{Y}
\]
system \eqref{4.21} is transformed to
\begin{equation}
I_{9}\hat{Y}' = (\lambda\tilde{T}^{-1}W\tilde{T} + D)Y, \label{4.24}
\end{equation}
where $D = \tilde{T}^{-1}Q_{1}\tilde{T} - \tilde{T}^{-1}\tilde{T}'$ ($Q_{1}$ is the part of the matrix $Q$ without the parameter $\lambda$).
We write down only two elements of the matrix $D$ which will be needed later on:
\begin{multline}
\begin{aligned}
d_{11} =& -\frac{1}{\sqrt{\hat{Z}\hat{\alpha}_{2}}}(i\omega(\hat{u} + \hat{\alpha}_{12}\sqrt{\frac{\hat{Z}}{\hat{\alpha}_{2}}}) + \frac{1}{2}\hat{Z}'\sqrt{\frac{\hat{\alpha}_{2}}{\hat{Z}}} + \frac{R_{43}\hat{\alpha}_{12}}{\hat{\alpha}_{2}} + \frac{R_{44}}{2} + \frac{\hat{\alpha}_{2}}{2}\left(\sqrt{\frac{\hat{Z}}{\hat{\alpha}_{2}}}\right)') -\\
&- \frac{1}{2}\hat{\alpha}_{12}Pr\hat{\alpha}_{0}\sqrt{\frac{\hat{\alpha}_{2}}{\hat{Z}}} - \frac{\sigma_{m}}{2b_{m}}(1 + \hat{\lambda})^{2}\frac{1}{\sqrt{\hat{Z}\hat{\alpha}_{2}}},\\
d_{22} =& -\frac{1}{\sqrt{\hat{Z}\hat{\alpha}_{2}}}(i\omega(\hat{u} + \hat{\alpha}_{12}\sqrt{\frac{\hat{Z}}{\hat{\alpha}_{2}}}) + \frac{1}{2}\hat{Z}'\sqrt{\frac{\hat{\alpha}_{2}}{\hat{Z}}} - \frac{R_{43}\hat{\alpha}_{12}}{\hat{\alpha}_{2}} - \frac{R_{44}}{2} + \frac{\hat{\alpha}_{2}}{2}\left(\sqrt{\frac{\hat{Z}}{\hat{\alpha}_{2}}}\right)') -\\
&- \frac{1}{2}\hat{\alpha}_{12}Pr\hat{\alpha}_{0}\sqrt{\frac{\hat{\alpha}_{2}}{\hat{Z}}} - \frac{\sigma_{m}}{2b_{m}}(1 + \hat{\lambda})^{2}\frac{1}{\sqrt{\hat{Z}\hat{\alpha}_{2}}}.
\end{aligned}\label{4.25}
\end{multline}

Let us get an asymptotic representation for the fundamental matrix of  system \eqref{4.24}. For this purpose we split the matrix $D$ into blocks according to representation \eqref{4.23}: the first diagonal block corresponds to nonzero diagonal elements  $-\frac{1}{\sqrt{\hat{Z}\hat{\alpha}_{2}}}$ and $\frac{1}{\sqrt{\hat{Z}\hat{\alpha}_{2}}}$, and the second diagonal block, on the contrary, corresponds to zero elements. Then, system \eqref{4.24} can be written in a more convenient form:
\begin{equation}
I_{10}\hat{Y}' = I_{10}\begin{pmatrix} \hat{Y}_{I}\\ \hat{Y}_{II}\end{pmatrix}' = (\lambda J + D)\begin{pmatrix} \hat{Y}_{I}\\ \hat{Y}_{II}\end{pmatrix}, \label{4.26}
\end{equation}
where
\begin{equation}
D = \begin{pmatrix}
D_{I}^{I} & D_{II}^{I}\\
D_{I}^{II} & D_{II}^{II}
\end{pmatrix}. \label{4.27}
\end{equation}
Using representation \eqref{4.27} of the matrix $D$, for the vector $\hat{Y}_{II}$ we get the following system:
\begin{equation}
(\hat{Y}_{II})' = D_{II}^{II}\hat{Y}_{II} + D_{I}^{II}\hat{Y}_{I}. \label{4.28}
\end{equation}
Assuming that the vector $\hat{Y}_{I}$ is known, we can now write down the system of fundamental solutions for the corresponding homogeneous system
\[
Y_{II}^{i}\bigg|_{y = -\frac{1}{2}} = \begin{pmatrix}0\\ \vdots \\ 0 \\ 1 \\0 \\ \vdots \\ 0 \end{pmatrix} -i\mbox{-th component}
\]
and the general solution of system \eqref{4.28}
\begin{equation}
\hat{Y}_{II} = \sum_{i = 3}^{10}C_{i}Y_{II}^{i} + \int_{-\frac{1}{2}}^{y}Y(y)Y^{-1}(s)D_{I}^{II}\hat{Y}_{I}, \label{4.29}
\end{equation}
where $Y(y)$ is the fundamental matrix consisting of the vectors $Y_{II}^{i}$.

Due to representation \eqref{4.29},  system for the remaining component $\tilde{Y}_{I}$ can be written as
\begin{multline}
\hat{Y}_{I}' = \lambda\begin{pmatrix} -\frac{1}{\sqrt{\hat{Z}\hat{\alpha}_{2}}} & 0\\ 0 & \frac{1}{\sqrt{\hat{Z}\hat{\alpha}_{2}}}\end{pmatrix}\hat{Y}_{I} + D_{I}^{I}\hat{Y}_{I} + D_{II}^{I}(\sum_{i = 3}^{10}C_{i}Y_{II}^{i}) +\\
+ D_{II}^{I}\int_{-\frac{1}{2}}^{y}Y(y)Y^{-1}(s)D_{I}^{II}\hat{Y}_{I}(y)ds. \label{4.30}
\end{multline}
Considering $D_{II}^{I}(\sum_{i = 3}^{10}C_{i}Y_{II}^{i})$ as a free term, let us find the fundamental matrix for  equation \eqref{4.30} in the following form:
\begin{equation}
\hat{Y} = (P_{0}(y) + \frac{1}{\lambda}P_{1}(y) + \frac{1}{\lambda^{2}}P_{2}(y) + \dots)\delta_{ij}e^{\lambda\Gamma_{j}(y)} + \frac{M_{0}}{\lambda^{2}} + \frac{M_{1}}{\lambda^{3}} + \dots, \label{4.31}
\end{equation}
where $\delta_{ij}$ is the Kronecker symbol, $\Gamma_{1}(y) = e^{-\int_{-\frac{1}{2}}^{y}\frac{1}{\sqrt{\hat{Z}\hat{\alpha}_{2}}}d\xi}$, $\Gamma_{2}(y) = e^{\int_{-\frac{1}{2}}^{y}\frac{1}{\sqrt{\hat{Z}\hat{\alpha}_{2}}}d\xi}$.
\begin{remark}
The first term in  formula \eqref{4.31} is the representation obtained by Birkhoff \cite{17} for the asymptotic solution of the differential equation
\begin{equation}
\tilde{Y}_{I}' = \lambda\begin{pmatrix} -\frac{1}{\sqrt{\hat{Z}\hat{\alpha}_{2}}} & 0\\ 0 & \frac{1}{\sqrt{\hat{Z}\hat{\alpha}_{2}}}\end{pmatrix}\tilde{Y}_{I} + D_{I}^{I}\hat{Y}_{I}. \label{4.32}
\end{equation}
\end{remark}
Substituting the matrix $\tilde{Y}$ into equation \eqref{4.30}, we get the following equation:
\begin{multline}
(P_{0}'(y) + \frac{1}{\lambda}P_{1}'(y) + \frac{1}{\lambda^{2}}P_{2}'(y) + \dots)\delta_{ij}e^{\lambda\Gamma_{j}(y)} + \lambda(P_{0}(y) + \frac{1}{\lambda}P_{1}(y) + \frac{1}{\lambda^{2}}P_{2}(y) + \dots)\\
\times \Lambda\delta_{ij}e^{\lambda\Gamma_{j}(y)} + \frac{M_{0}'}{\lambda^{2}} + \frac{M_{1}'}{\lambda^{3}} + \dots =\\
= \lambda\Lambda(P_{0}(y) + \frac{1}{\lambda}P_{1}(y) + \frac{1}{\lambda^{2}}P_{2}(y) + \dots)\delta_{ij}e^{\lambda\Gamma_{j}(y)} + \lambda\Lambda(\frac{M_{0}}{\lambda^{2}} + \frac{M_{1}}{\lambda^{3}} + \dots) +\\
+ D_{I}^{I}(P_{0}(y) + \frac{1}{\lambda}P_{1}(y) + \frac{1}{\lambda^{2}}P_{2}(y) + \dots)\delta_{ij}e^{\lambda\Gamma_{j}(y)} + D_{I}^{I}(\frac{M_{0}}{\lambda^{2}} + \frac{M_{1}}{\lambda^{3}} + \dots) +\\
+ D_{II}^{I}\int_{-\frac{1}{2}}^{y}Y(y)Y^{-1}(s)D_{I}^{II}\bigg[(P_{0}(s) + \frac{1}{\lambda}P_{1}(s) + \frac{1}{\lambda^{2}}P_{2}(s) + \dots)\delta_{ij}e^{\lambda\Gamma_{j}(s)} +\\
+ \frac{M_{0}(s)}{\lambda^{2}} + \frac{M_{1}(s)}{\lambda^{3}} + \dots \bigg]ds, \quad \Lambda = diag\{-\frac{1}{\sqrt{\hat{Z}\hat{\alpha}_{2}}}, \frac{1}{\sqrt{\hat{Z}\hat{\alpha}_{2}}}\}. \label{4.33}
\end{multline}
Comparing coefficients by the same powers of $\lambda$ in terms consisting of the matrix $\delta_{ij}e^{\lambda\Gamma_{j}(y)}$ as well as in terms free from it and integrating by parts a required number of times, we get:
\begin{equation}
P_{0}\Lambda = \Lambda P_{0}, \label{4.34}
\end{equation}
\begin{equation}
P_{0}' + P_{1}\Lambda = \Lambda P_{1} + D_{I}^{I}P_{0}, \label{4.35}
\end{equation}
\begin{equation}
P_{1}' + P_{2}\Lambda = \Lambda P_{2} + D_{I}^{I}P_{1} + D_{II}^{I}D_{I}^{II}P_{0}\Lambda^{-1}, \label{4.36}
\end{equation}
\begin{equation}
\Lambda M_{0} = D_{I}^{II}Y D_{II}^{I}(-\frac{1}{2})\Lambda^{-1}(-\frac{1}{2}). \label{4.37}
\end{equation}
Then, from  equality \eqref{4.34} we necessarily have that $P_{0}(y)$ is a diagonal matrix,
\[
P_{0}(y) = \begin{pmatrix}
p_{1}(y) & 0\\
0 & p_{2}(y)
\end{pmatrix},
\]
and equality \eqref{4.35} for the case of diagonal elements gives us the Cauchy problem
\begin{equation}
p_{i}' = d_{ii}^{I}p_{i}, \quad p_{i}(-\frac{1}{2}) = 1, \quad i = 1,2, \label{4.38}
\end{equation}
where $d_{ii}^{I}$ are the diagonal elements of the matrix $D$ (see formulas \eqref{4.25}).

Solving problems \eqref{4.38}, we find functions $p_{i}(y)$:
\[
p_{i}(y) = e^{-\int_{\frac{-1}{2}}^{y}d_{ii}^{I}(\xi)d\xi}, \, i = 1,2
\]
(below the top index $I$ is omitted).

Now equality \eqref{4.35} enable us to define off-diagonal elements of the matrix $P_{1}(y)$. At the same time, diagonal elements are found from equality \eqref{4.36}. By induction we can find all the matrices $P_{i}(y)$, $i = 1,2, \dots$. Equalities similar to \eqref{4.37} enable one to define the matrices $M_{i}$, $i = 0,1,\dots$.

The presence of the free term $D_{II}^{I}(\sum_{i = 3}^{10}C_{i}Y_{II}^{i})$ due to the method of variation of constants leads to the appearance of additional terms in representation \eqref{4.31} which contain powers $\frac{1}{\lambda}$ (i.e., $\frac{1}{\lambda^{k}}$, $k = 1,\dots$) as multipliers. As will see below such terms do not influence on the main term of the asymptotic representation of the spectrum and, so, in representation \eqref{4.31} we can consider only the main term:
\begin{equation}
\hat{Y} = P_{0}(y)\delta_{ij}e^{\lambda\Gamma_{j}(y)}. \label{4.39}
\end{equation}
Remembering about the fundamental matrix of equation \eqref{4.28}, together with equation \eqref{4.39} we obtain the main term of the asymptotic representation of the fundamental matrix $W_{\tilde{Y}}$ for  system \eqref{4.24}:
\begin{equation}
W_{\hat{Y}} = \begin{pmatrix}
e^{-\lambda\int_{-\frac{1}{2}}^{y}\frac{1}{\sqrt{\hat{Z}\hat{\alpha}_{2}}}d\xi}p_{1}(y) & 0 & 0 & \dots & 0\\
0 & e^{-\lambda\int_{-\frac{1}{2}}^{y}\frac{1}{\sqrt{\hat{Z}\hat{\alpha}_{2}}}d\xi}p_{2}(y) & 0 & \dots & 0\\
0 & 0 & y_{3}^{3} & \dots & y_{10}^{3}\\
\vdots & \vdots & \vdots & \vdots & \vdots\\
0 & 0 & y^{10}_{3} & \dots &  y_{10}^{10}
\end{pmatrix}, \label{4.40}
\end{equation}
where $y_{i}^{j}$, $i,j = 3,\dots,10$ are components of the fundamental system for \eqref{4.28} composed from columns of the matrix $Y(y)$.

\begin{remark}
In this work we do not give a justification that representation \eqref{4.31} is indeed an asymptotic series as well as the described representation for the fundamental matrix  $W_{\hat{Y}}$ (or, more precisely, for its ``complete'' version) is also the subjectfor future research. We only note that such a fact was established  by Birkhoff \cite{17} for equation \eqref{4.30} (when in the right-hand side of \eqref{4.30} there are no terms with the coefficients $M_{0}$, $M_{1}$,\dots, i.e., there is no part of series corresponding to the integral term of equation \eqref{4.30}). In this case in each of the half-planes Re$\lambda > 0$ and Re$\lambda < 0$ asymptotic series are different from each other \cite{19}. By the way, considering the integral term as the free one  and using the method of variation of constants, we can  find the matrices $M_{0}$, $M_{1}$,\dots  by another way  (the idea of finding the fundamental matrix by using the method of variation of constants is described in \cite{18}).
\end{remark}

Let us remember the boundary conditions \eqref{4.18}, \eqref{4.19} which after the change of unknown
\[
\tilde{Y} = \tilde{T}\hat{Y}
\]
are transformed to
\begin{multline}
y = -\frac{1}{2}: \quad \hat{y}_{1} = \hat{y}_{2}, \, \hat{y}_{6} = 0, \, (1 + \hat{\lambda})\sqrt{\frac{\hat{Z}}{\hat{\alpha}_{2}}}\hat{y}_{1} + (1 + \hat{\lambda})\sqrt{\frac{\hat{Z}}{\hat{\alpha}_{2}}}\hat{y}_{2} + b_{m}\hat{y}_{7} = 0,\\
i\omega(1 + \bar{\theta})\hat{y}_{1} + i\omega(1 + \bar{\theta})\hat{y}_{2} - y_{4} = 0; \label{4.41}
\end{multline}
\begin{multline}
y = \frac{1}{2}: \quad \hat{y}_{1} = \hat{y}_{2}, \, \hat{y}_{6} = 0, \, (1 + \hat{\lambda})\sqrt{\frac{\hat{Z}}{\hat{\alpha}_{2}}}\hat{y}_{1} + (1 + \hat{\lambda})\sqrt{\frac{\hat{Z}}{\hat{\alpha}_{2}}}\hat{y}_{2} + b_{m}\hat{y}_{7} = 0,\\
i\omega\hat{y}_{1} + i\omega\hat{y}_{2} - y_{4} = 0. \label{4.42}
\end{multline}
In view of representation \eqref{4.40}, they can be written as follows:
\begin{equation}
\det\begin{pmatrix}
L\\
RW_{\hat{Y}}(\frac{1}{2})
\end{pmatrix} = 0, \label{4.43}
\end{equation}
where
\[
L = \begin{pmatrix}
1 & -1 & 0 & 0 & 0 & 0 & 0 & 0 & 0 & 0\\
i\omega(1 + \bar{\theta}) & i\omega(1 + \bar{\theta}) & 0 & -1 & 0 & 0 & 0 & 0 & 0 & 0\\
0 & 1 & 0 & 0 & 0 & 0 & 0 & 0 & 0 & 0\\
0 & 0 & 0 & 0 & 0 & 1 & 0 & 0 & 0 & 0\\
0 & 0 & 0 & 0 & 0 & 0 & 0 & 0 & 0 & 1\\
\end{pmatrix},
\]
\[
R = \begin{pmatrix}
1 & -1 & 0 & 0 & 0 & 0 & 0 & 0 & 0 & 0\\
i\omega & i\omega & 0 & -1 & 0 & 0 & 0 & 0 & 0 & 0\\
0 & 1 & 0 & 0 & 0 & 0 & 0 & 0 & 0 & 0\\
0 & 0 & 0 & 0 & 0 & 1 & 0 & 0 & 0 & 0\\
0 & 0 & 0 & 0 & 0 & 0 & 0 & 0 & 0 & 1\\
\end{pmatrix}.
\]
After elementary transforms of the determinant and the use of Laplace theorem on decomposition of a determinant into a product of minors, it turns out that equation \eqref{4.43} is equivalent to the following equation:
\[
e^{\lambda\int_{-\frac{1}{2}}^{\frac{1}{2}}\frac{1}{\sqrt{\hat{Z}\hat{\alpha}_{2}}}d\xi}p_{2}(1) - e^{-\lambda\int_{-\frac{1}{2}}^{\frac{1}{2}}\frac{1}{\sqrt{\hat{Z}\hat{\alpha}_{2}}}d\xi}p_{1}(1) = 0
\]
or
\begin{equation}
e^{\lambda\int_{-\frac{1}{2}}^{\frac{1}{2}}\frac{1}{\sqrt{\hat{Z}\hat{\alpha}_{2}}}d\xi}e^{\int_{-\frac{1}{2}}^{\frac{1}{2}}d_{11}(\xi)d\xi} - e^{-\lambda\int_{-\frac{1}{2}}^{\frac{1}{2}}\frac{1}{\sqrt{\hat{Z}\hat{\alpha}_{2}}}d\xi}e^{\int_{-\frac{1}{2}}^{\frac{1}{2}}d_{22}(\xi)d\xi} = 0. \label{4.44}
\end{equation}
Recalling formulas \eqref{4.25}, we get the spectrum representation:
\begin{multline}
\lambda_{k} = \left[\int_{-\frac{1}{2}}^{\frac{1}{2}}\frac{1}{\sqrt{\hat{Z}\hat{\alpha}_{2}}}d\xi\right]^{-1}\left(\int_{-\frac{1}{2}}^{\frac{1}{2}}\frac{1}{\sqrt{\hat{Z}\hat{\alpha}_{2}}}(i\omega\hat{u} + \frac{R_{43}\hat{\alpha}_{12}}{\hat{\alpha}_{2}} + \frac{R_{44}}{2}d\xi) + k\pi i\right)+\\
+ O(\frac{1}{k}), \, k \to \infty, \label{4.45}
\end{multline}
where $O (\cdot )$ is again the usual big $O$ notation. The proof of Theorem 1 is complete.

\begin{remark}
Using representation \eqref{4.31}, we can get an asymptotic representation of $\lambda_{k}$ with any order of accuracy defined by the powers of $\frac{1}{k}$ (see also \cite{17}).
\end{remark}

Now, as a consequence, we get the following result. If the Poiseuille-type flow described in Sect. 2 is asymptotically stable by Lyapunov, then the following inequality it is necessarily holds:
\begin{equation}
\mbox{Re}\lambda_{k} = \left[\int_{-\frac{1}{2}}^{\frac{1}{2}}\frac{1}{\sqrt{\hat{Z}\hat{\alpha}_{2}}}d\xi\right]^{-1}\left(\int_{-\frac{1}{2}}^{\frac{1}{2}}\frac{1}{\sqrt{\hat{Z}\hat{\alpha}_{2}}}(\frac{R_{43}\hat{\alpha}_{12}}{\hat{\alpha}_{2}} +\frac{R_{44}}{2})d\xi\right) < 0. \label{4.46}
\end{equation}
By virtue of the formulas for $R_{43}$ and $R_{44}$, inequality \eqref{4.46} can be rewritten as
\begin{equation}
\mbox{Re}\lambda_{k} = \left[\int_{-\frac{1}{2}}^{\frac{1}{2}}\frac{1}{\sqrt{\hat{Z}\hat{\alpha}_{2}}}d\xi\right]^{-1}\left(\int_{-\frac{1}{2}}^{\frac{1}{2}}\frac{\hat{\chi}_{0}^{*}}{\sqrt{\hat{Z}\hat{\alpha}_{2}}}(\hat{a}_{11}(\frac{\bar{k}}{3} + \beta) + \frac{1}{2W})d\xi\right) < 0, \label{4.47}
\end{equation}
and inequality \eqref{4.47} implies inequality \eqref{3.10}. The proof of Theorem 2 is complete.

The authors are grateful to A.V. Yegitov for his help in the preparation of the manuscript of this paper.

This work was supported by the RFBR grant 17-01-00791a.

\leftline{For more information, please visit the journal website:}

\centerline{http://www.worldscinet.com/jhde.html}


\begin{thebibliography}{0}

\bibitem{20}
V.N. Pokrovskii,
{\it The mesoscopic theory of polymer dynamics, 2nd Ed. / V.N. Pokrovskii},
(Springer, Dordrecht-Heidelberg, London - New-York, 2010).

\bibitem{21}
G.V. Pishnograi, V.N. Pokrovski, Y.G. Yanovski, I.F. Obraztsov, Y.N. Karnet,
Defining equation for nonlinear viscoelastic (polymeric) mediums in zeroth approximation by parameters of molecular theory and consequances for shear and elongation,
{\it DAN},
{\bf 355(9)} (1994),
612--615 (in Russian).

\bibitem{1}
Yu.A. Altukhov, A.S. Gusev, G.V. Pishongrai,
{\it Introduction into mesoscopic theory of flowing polymeric systems},
(Alt. GPA, Barnaul, 2012) (in Russian).

\bibitem{22}
K.B. Koshelev, G.V. Pishnograi, A.Ye. Kuznetsov, M.Yu. Tolstikh,
Dependancy of hydrodynamic characteristics of the polymer melts flow in converging channel from temperature,
{\it Mechanics of composite materials and constructions},
{\bf 22(2)} (2016),
175--191 (in Russian).

\bibitem{23}
A.N. Krylov,
On the stability of a Poiseuille flow in a planar channel,
{\it DAN},
{\bf 158(5)} (1964),
978--981 (in Russian).

\bibitem{24}
W. Heisenberg,
Uber Stabilitat und Turbulenz von Flussingkeitsstromen
{\it Ann. Phys.},
{\bf 74} (1924),
577--627.

\bibitem{25}
E. Grenier, Y. Guo, T.T. Nguyen,
Spectral instability of characteristic boundary layer flows,
{\it Duke Math J.},
{\bf 165(16)} (2016),
 3085--3146.

\bibitem{2}
L.I. Sedov,
{\it A Course in Continuum Mechanics: Basic Equations and Analytical Techniques (Volume 1)}
(Wolters-Noordhoff Publishing, 1971).

\bibitem{3}
L.G. Loitsynski,
{\it Mechanics of Liquids and Gases},
(BHB, 1995).

\bibitem{4}
A.B. Vatazhin, G.A. Lubimov, S.A. Regirer,
{\it Magnetohydrodynamic flows in channels}
(Nauka, Moscow, 1970) (in Russian).

\bibitem{5}
Shih-i Pai,
{\it Introduction to the Theory of Compressible Flow},
(D. Van Nostrand Co, Princeton, 1962).

\bibitem{6}
A.M. Blokhin, A.S. Rudometova,
Stationary solutions of the equations for nonisothermal electroconvection of a weakly conducting incompressible polymeric liquid,
{\it Journal of Applied and Industrial Mathematics},
{\bf 9(2)} (2015),
147--156.

\bibitem{7}
Y. Shibata,
On the R-Boundedness for the Two Phase Problem with Phase Transition: Compressible - Incompressible Model Problem,
{\it Funkcialay Ekvacioj},
{\bf 59} (2016),
 243--287.

\bibitem{8}
N.A. Slezkin,
{\it Dynamics of a Viscous Incompressible Fluid},
(Gos. tech. teor. izdat., Moscow, 1955) (in Russian).

\bibitem{9}
A.N. Akhiezer, N.A. Akhiezer,
{\it Electromagnetism and electromagnetic waves Ёлектромагнетизм и электромагнитные волны}
(Higher School, Moscow, 1985) (in Russian).

\bibitem{10}
K. Nordling, J. Osterman,
{\it Physics Handbook for Science and Engineering},
(Chartwell-Bratt, 1996).

\bibitem{11}
L.D. Landau, Ye.M. Lifshitz,
{\it Electrodynamics of continuum media},
(Pergamon Press, 1960).

\bibitem{12}
A.M. Blokhin, A.V. Yegitov, D.L. Tkachev,
Asymptotics of the Spectrum of a Linearized Problem of the Stability of a Stationary Flow of an Incompressible Polymer Fluid with a Space Charge,
{\it Computational Mathematics and Mathematical Physics},
{\bf 56(1)} (2018),
102--117.

\bibitem{13}
 Alexander Blokhin, Dmitry Tkachev and Aleksey Yegitov,
Spectral asymptotics of a linearized problem for an incompressible weakly conducting polymeric fluid,
{\it ZAMM (Z. Angrew. Math. Mech.)},
{\bf 98(4)} (2018),
 589--601.

\bibitem{14}
A.M. Blokhin, D.L. Tkachev,
Analogue of the Poiseuille flow for incompressible polymeric fluid with volume charge. Asymptotics of the linearized problem spectrum,
{\it IOP Conf. Series: Journal of Physics: Conf. Series},
{\bf 894 (012096)} (2017),
1--6.

\bibitem{BS}
A.M. Blokhin, R.Y. Semenko, Stationary magnetohydrodynamic flows of a non-isothermal incompressible polymeric liquid in the flat channel,
{\it Bulletin of the South Ural State University, Ser. Mathematics Modeling, Programming \& Computer Software},
{\bf 11(4)}, 2018,
41--54.

\bibitem{15}
A.M. Blokhin, A.V. Yegitov, D.L. Tkachev,
Linear instability of solutions in a mathematical model describing polymer flows in an infinite channel,
{\it Computational Mathematics and Mathematical Physics},
{\bf 55(5)} (2015),
 848--873.

\bibitem{16}
A.M. Blokhin, D.L. Tkachev,
Linear asymptotic instability of a stationary flow of a polymeric medium in a plane channel in the case of periodic perturbations,
{\it Journal of Applied and Industrial Mathematics},
{\bf 8(4)} (2014),
467--478.

\bibitem{17}
G.D. Birkhoff,
{\it Collected mathematical papers},
(AMS,  New York,1950).

\bibitem{19}
K.V. Brushlinski,
On growth of mixed problem solution in case of incomplete eigen-functions,
{\it Izvestiya AN SSSR, seriya matematika},
{\bf 23} (1959),
 893--912 (in Russian).

\bibitem{18}
M.V. Fedoruk,
{\it Asymptotic methods for ordinary differential equations},
(Nauka, Moscow, 1983) (in Russian).

\end{thebibliography}
\end{document}